\documentclass[12pt]{article}

\usepackage[english]{babel}
\usepackage{setspace}
\usepackage[letterpaper]{geometry}
\usepackage{graphicx}
\usepackage[colorlinks=true, allcolors=blue]{hyperref}
\usepackage[comma,authoryear,round]{natbib}
\usepackage{amsmath,amsthm,amsfonts,amssymb}
\usepackage{cases}
\usepackage{nccmath,stmaryrd}
\usepackage{nicefrac}
\usepackage{dsfont}
\usepackage{commath}
\usepackage{mathtools}
\usepackage[mathscr]{euscript}
\usepackage{accents}
\usepackage{wrapfig}
\usepackage{enumitem}
\usepackage[capitalise]{cleveref} 
\usepackage{mathtools} 
\usepackage{color,xcolor}
\usepackage{comment}

\newcommand{\blind}{0}

\newtheorem{theorem}{Theorem}[section]
\theoremstyle{definition}
\newtheorem{definition}{Definition}[section]
\theoremstyle{plain}
\newtheorem{lemma}{Lemma}[section]

\newcommand{\Like}{\operatorname{L}}
\newcommand{\Spec}{\operatorname{S}}
\newcommand{\Res}{\operatorname{R}}
\newcommand{\Ng}{\operatorname{NG}}

\renewcommand{\Re}{\mathbb{R}}

\def\spacingset#1{\renewcommand{\baselinestretch}%
{#1}\small\normalsize} \spacingset{1}

\begin{document}

\title{\bf On inference for modularity statistics in structured networks}

\date{}

\if0\blind
{
  \author{Anirban Mitra, Konasale Prasad, Joshua Cape\thanks{
    The authors thank Nicholas Theis and the entire CONCEPT lab for real data expertise. This research uses data from the UK Biobank, a major biomedical database, obtained from the UK Biobank Resource under application number 68923 (PI: Konasale Prasad). This research was supported in part by the University of Pittsburgh Center for Research Computing through the resources provided. Specifically, this work used the H2P cluster which is supported by NSF award number OAC-2117681. JC gratefully acknowledges support from the University of Wisconsin--Madison, Office of the Vice Chancellor for Research and Graduate Education, with funding from the Wisconsin Alumni Research Foundation.} \\ \\
    Department of Statistics, University of Pittsburgh \\
    Department of Psychiatry, University of Pittsburgh \\
    Department of Statistics, University of Wisconsin--Madison}
} \fi

\if1\blind
{
\title{\bf On inference for modularity statistics in structured networks}
}\fi

\maketitle

\bigskip
\begin{abstract}
    This paper revisits the classical concept of network modularity and its spectral relaxations used throughout graph data analysis. We formulate and study several modularity statistic variants for which we establish asymptotic distributional results in the large-network limit for networks exhibiting nodal community structure. Our work facilitates testing for network differences and can be used in conjunction with existing theoretical guarantees for stochastic blockmodel random graphs. Our results are enabled by recent advances in the study of low-rank truncations of large network adjacency matrices. We provide confirmatory simulation studies and real data analysis pertaining to the network neuroscience study of psychosis, specifically schizophrenia. Collectively, this paper contributes to the limited existing literature to date on statistical inference for modularity-based network analysis. Supplemental materials for this article are available online.
\end{abstract}

\vfill

\newpage
\spacingset{1.5}

\section{Introduction}
\label{sec:introduction}

Networks and graph-structured data, comprised of nodes (i.e.,~entities) and edges (i.e.,~pairwise interactions), are frequently analyzed in an effort to determine the extent to which nodal community structure is present. Modularity-based measures have emerged as popular criteria for quantifying the existence of nodal organization via edge connectivity patterns, commonly described as \emph{modules}, \emph{clusters}, or \emph{communities} \citep{newman2018networks}. Examples of so-called modules or groups of network nodes include friendship circles among people in social networks \citep{zachary1977aninformation,leung2014interactive,fitzpatrick2018theory,tianxi2020network} and ensembles of brain regions exhibiting high functional or structural connectivity in neuroimaging scans \citep{rubinov2010complex,power2011functional,grayson2017development}.

Finding modules has meaningful downstream consequences for subsequent interventions and inference, for example, towards promoting diversity in social or political interactions \citep{eulau1981social,porter2005network}, and towards developing precision medicine or therapies for health disorders \citep{capriotti2019integrating,tan2019network}. Modularity analysis has been applied in the study of mobile communication networks to suggest that linguistic `borders' or `divides' are influenced by a combination of geographic, political, and demographic factors \citep{blondel2010regions}. Elsewhere, \cite{guimera2005worldwide} applies modularity maximization to detect communities in the worldwide air transportation network, finding that module structure and formation are driven by geopolitics and not simply by geographic physical proximity alone.

In practice, given a collection of observed networks, analysts often compute various graph-theoretic summary statistics, including modularity values, for the purpose of comparing and contrasting the networks. Differences between individual networks or populations of networks are then reported and interpreted, sometimes but not always accompanied by empirical, simulation-based, or model-assisted proxies for quantifying uncertainty in the reported statistics. For example, colloquially, Network~A or (networks in) Population~A might be reported as being `more modular' (or, say, more connected, or sparser) than Network~B or (networks in) Population~B.

The purpose of this article is to contribute towards a deeper theoretical and methodological understanding of modularity statistics and their variability in the context of random graph models. The limited availability of statistical inference tools for modularity-based analysis has been noted elsewhere in the literature, notably in \cite{ma2020theasymptotic} which, pertinent to the present paper, establishes the asymptotic distribution of a modularity statistic for large data matrices arising from the Gaussian Orthogonal Ensemble. Other relevant works include \cite{li2020asymptotic} which establishes asymptotic normality of the Newman--Girvan modularity statistic for large networks under the null hypothesis of so-called free labeling, for the purpose of testing whether a given partition has arisen completely at random. \cite{zhang2017hypothesis} provides a hypothesis testing framework for assessing the significance of detected communities where the space of null models is the set of simple graphs having the same degree sequence as the observed graph. See also the highly influential asymptotic analysis carried out in \cite{bickel2009nonparametric}.

Here, we briefly preface the content that follows. In this paper, we focus on several well-motivated cases of a to-be-specified general modularity function, drawing on the spectral (eigen) properties of large network adjacency matrices. For regimes where ground truth community assignments are asymptotically uniformly perfectly recoverable in stochastic blockmodel random graphs, we establish the asymptotic normality of modularity values with explicit asymptotic bias and variance expressions. We provide detailed simulations to illustrate these findings and provide real data examples involving human brain networks. Further extensions and complements are also discussed.

\section{Background}
\label{sec:background}

\subsection{Random graph models}
\label{sec:random_graph_models}

A diversity of random graph models are encountered throughout the network-focused literature in statistics, mathematics, and physics. Two extremal examples at opposite ends of the proverbial spectrum include homogeneous and inhomogeneous Erd\H{o}s--R\'{e}nyi graphs \citep{erdos59random,gilbert1959random,bollobas2007phase}. In this paper, we compromise between simplicity and generality by focusing on stochastic blockmodel (SBM) graphs \citep{holland1983stochastic}.

\begin{definition}[Stochastic blockmodel random graphs]\label{def_sbm}
    Let $K \ge 2$ be a positive integer. Let $\boldsymbol{\pi} \in \mathcal{S}_{K-1}$ be a non-negative vector in the interior of the $(K-1)$-dimensional simplex in $\Re^{K}$, whence $0<\pi_{k}<1$ for each $1 \le k \le K$ with $\sum_{k=1}^{K}\pi_{k} = 1$. Let $\mathbf{B} \in (0,1)^{K \times K}$ be a symmetric matrix of probabilities. First, sample i.i.d. categorical random variables $\tau_{1}, \dots, \tau_{n}$ according to $\mathbb{P}[\tau_{i} = k] = \pi_{k}$ to form the membership vector $\boldsymbol{\tau} = (\tau_{1},\dots,\tau_{n})^{\top}$. Second, conditional on $\boldsymbol{\tau}$, generate a binary symmetric random adjacency matrix $\mathbf{A} \in \{0,1\}^{n \times n}$ according to $A_{ij} \sim \operatorname{Bernoulli}(\rho_{n}B_{\tau_{i}\tau_{j}})$ independently for all $i \le j$, where $\rho_{n} \in (0,1]$ denotes a sparsity factor. We then write $(\mathbf{A}, \boldsymbol{\tau}) \sim \operatorname{SBM}(\mathbf{B}, \boldsymbol{\pi})$ with accompanying sparsity factor $\rho_{n}$. We write $\mathbf{A} \sim \operatorname{SBM}(\mathbf{B}, \boldsymbol{\pi})$ with sparsity factor $\rho_{n}$ when only $\mathbf{A}$ is observed, i.e.,~when $\boldsymbol{\tau}$ is integrated out from $(\mathbf{A}, \boldsymbol{\tau})$.
\end{definition}

Stochastic blockmodeling provides a useful framework for modeling nodal clustering and connectivity. The block structure in SBMs corresponds to the concept of nodal communities. Further, SBMs need not place any additional restrictions \emph{a priori} on the within-block versus between-block edge connectivity probabilities. For example, letting $\mathbb{I}_{\{\cdot\}}$ denote the binary indicator function, an SBM graph with edge probabilities given by $B_{\tau_{i}\tau_{j}} = a \cdot \mathbb{I}_{\{\tau_{i} = \tau_{j}\}} + b \cdot \mathbb{I}_{\{\tau_{i} \neq \tau_{j}\}}$ for all $1 \le \tau_{i}, \tau_{j} \le K$ exhibits assortativity (i.e.,~homophily) when $a > b$ but is disassortative (i.e.,~heterophilic) when $a < b$. Beyond homogeneous assortativity and disassortativity, more complex connectivity phenomena are enabled by the at most $K(K+1)/2$ distinct entries of $\mathbf{B}$, such as core-periphery network structure.

Stochastic blockmodels have received considerable research attention in recent years \citep{abbe2018community,lee2019review}. As such, they produce random graphs whose statistical properties are increasingly better and better understood. While outside the scope of the present paper, popular SBM variants include degree-corrected SBMs \citep{karrer2011stochastic} which permit node-specific degree heterogeneity, mixed-membership SBMs \citep{airoldi2008mixed} which permit node-specific weighted multi-block memberships, weighted SBMs \citep{aicher2013adaptingARXIV,aicher2014learning} which permit heterogeneous edge weights, popularity-adjusted SBMs \citep{sengupta2018blockmodel,koo2023popularity}, and covariate-adjusted SBMs \citep{sweet2015incorporating,zhang2019node}.

\subsection{Basics of network modularity}
\label{sec:basics_modularity}

Let $\mathbf{A} = (A_{ij}) \in \Re^{n \times n}$ denote a (possibly weighted) adjacency matrix corresponding to an undirected $n$-node graph $\mathcal{G}$. Given a posited `null network' $\mathbf{P} = (P_{ij}) \in \Re^{n \times n}$ serving as a benchmark for $\mathbf{A}$, and given a partition of $\llbracket n \rrbracket \coloneqq \{1,2,\dots,n\}$ encoded by the cluster membership vector $\boldsymbol{\tau}$, a general formulation of the modularity quality function \citep{bazzi2016community} is given by
\begin{equation}
	\label{eq:modularity_generic}
    Q
    \equiv
    Q(\boldsymbol{\tau}; \mathbf{A}, \mathbf{P})
    \coloneqq
    \sum_{1 \le i,j \le n} (A_{ij} - P_{ij}) \mathbb{I}_{\{\tau_{i} = \tau_{j}\}}.
\end{equation}
In words, $Q$ reflects the partition-specific discrepancy between the total edge weights in the observed network and the total anticipated edge weights in the referential, posited null network. Large values of $Q$ provide partition-specific evidence vis-\'{a}-vis $\boldsymbol{\tau}$ that the observed network is more modular than by chance alone. Evaluating this discrepancy across numerous partitions $\boldsymbol{\tau} \in \mathcal{T}$, or attempts at all possible partitions, naturally leads to the modularity maximization problem
\begin{equation}
	\label{eq:modularity_maximization}
    (\arg)\;\underset{\boldsymbol{\tau} \in \mathcal{T}}{\max}\;Q(\boldsymbol{\tau}; \mathbf{A}, \mathbf{P})
\end{equation}
and computational schemes for solving relaxations thereof. The use of parentheses in \cref{eq:modularity_maximization} indicates interest both in the maximizing partition and the maximum modularity value.

For undirected unweighted graphs $\mathcal{G}$, historically the most commonly adopted choice of modularity function is due to Newman and Girvan, denoted by $Q_{\Ng}$, with the specification $P_{ij} = k_{i}k_{j}/(2m)$, where $k_{i} = \sum_{j=1}^{n}A_{ij}$ denotes the degree of node $i$ in $\mathbf{A}$, and $m = \sum_{i=1}^{n}k_{i}/2$ denotes the total number of graph edges \citep{newman2004finding}. Further dividing $Q_{\Ng}$ by $2m$ yields normalized modularity values in a bounded interval. The choice $Q_{\Ng}$ has the attractive property of specifying a data-dependent benchmark yet is not assumption-free, for it takes the configuration model as the underlying null model for observable random graphs \citep{newman2018networks}.

To date, numerous algorithms and optimization procedures have been proposed for \cref{eq:modularity_maximization} and specifically for discovering community structure using $Q_{\Ng}$. Agglomerative algorithms for modularity maximization include \cite{newman2004fast}, with \cite{clauset2004finding} adopting a fast greedy approach, and \cite{blondel2008fast} proposing the Louvain heuristic. \cite{brandes2008modularity} provides an integer linear programming approach for modularity maximization. \cite{traag2019fromlouvain} suggests the Leiden algorithm as an improvement over the Louvain algorithm by allowing for and ``aggregating over refined partitioning of the communities'' during the stage of ``local moving of nodes.'' Other methods for optimizing modularity include extremal optimization \citep{duch2005community} and simulated annealing \citep{reichardt2006statistical}.

Yet another popular approach for modularity maximization is to consider spectral relaxations via the eigendecomposition of the modularity matrix \citep{newman2006modularity}. In the case of two communities, defining the membership vector as having entries either $+1$ or $-1$ permits rewriting $Q_{\Ng}$ as a scaled quadratic form. The element-wise choice of sign is derived from the component-wise signs of the eigenvector corresponding to the largest positive eigenvalue of the modularity matrix. To generalize the method for detecting more than two communities, successive bipartitioning can be applied. \cite{newman2006finding} adopts an alternative approach for $K>2$ by considering an $n \times K$ matrix whose one-hot-encoded columns represent node-community correspondences.

We mention in passing that modularity considerations for graphons have recently been investigated in \cite{klimm2021modularity}, defined akin to Newman--Girvan modularity but with a modularity surface in place of a modularity matrix. The study of multi-layer networks has also recently garnered attention \citep{dong2012clustering,lei2019consistent,bhattacharyya2020generalARXIV} thereby motivating further extensions of modularity functions and modularity maximization concepts \citep{mucha2010community,bazzi2016community,dedomenico2015identifying,dedominico2015structural,dedominico2017multilayer}.

\section{Modularity-based  inference}
\label{sec:modularity_based_inference}

\subsection{Modularity variants}
\label{sec:modularity_variants}

As mentioned above, a common approach to modularity maximization and analysis is to resort to spectral relaxations of \cref{eq:modularity_maximization}, which, in the language of this paper, leads to computing eigenvalues and eigenvectors associated to the matrix $\mathbf{A} - \mathbf{P}$. We adopt a related but slightly different approach, motivated by several key observations in \cite{tang2022asymptotically} and related works. The first observation is that so-called strong consistency of eigenvector-based clustering holds for not-too-sparse stochastic blockmodel graphs \citep{xie2022entrywise,rubindelanchy2022astatistical, lyzinski2014perfect}. In particular, for such SBMs, $\widehat{\boldsymbol{\tau}}$ derived from the leading eigenvectors of $\mathbf{A}$ perfectly and uniformly recovers the unobserved true memberships encoded in $\boldsymbol{\tau}$ in $\ell_{\infty}$ vector norm with probability rapidly tending to one in the large-network limit. The second key observation is that computing truncated eigendecompositions of large data matrices exhibiting approximately low-rank population-level structure has the effect of data denoising. Quantitatively, a suitable low-rank truncation of $\mathbf{A}$, denoted by $\widehat{\mathbf{A}}$, yields a better approximation to $\mathbb{E}[\mathbf{A} \mid \boldsymbol{\tau}]$ than does $\mathbf{A}$ itself. Taken together, we are led to formulate and study the following three modularity statistics.

\begin{definition}[\bf Likelihood-based, Spectral-based and Residual-based modularities]
	\label{def_mods3}
	Denote the full eigendecomposition of the $n \times n$ symmetric matrix $\mathbf{A}$ by $\mathbf{A} = \sum_{i=1}^{n}\widehat{\lambda}_{i}\widehat{\mathbf{u}}_{i}\widehat{\mathbf{u}}_{i}^{\top}$ with eigenvalues $|\widehat{\lambda}_{1}| \ge \dots \ge |\widehat{\lambda}_{n}|$ and orthonormal eigenvectors $\widehat{\mathbf{u}}_{1},\dots,\widehat{\mathbf{u}}_{n} \in \Re^{n}$. We write $\widehat{\mathbf{A}} = \sum_{i=1}^{d} \widehat{\lambda}_{i}\widehat{\mathbf{u}}_{i}\widehat{\mathbf{u}}_{i}^{\top}$ to denote the top rank $d$ approximation of $\mathbf{A}$. In what follows, the value $d$ corresponds to an underlying population-level quantity (matrix rank) that is consistently estimable using $\mathbf{A}$ and will be subsequently discussed in greater detail. Given a specified cluster membership vector $\boldsymbol{\tau}$, we consider the following three modularity variants.
	\begin{subequations}
		\label{modularities}
	\begin{align}
	    Q_{\Like}
	    &\coloneqq
	    \sum_{i,j = 1}^{n} \left({A}_{ij} - {P}_{ij}\right) \mathbb{I}_{\{\tau_{i} = \tau_{j}\}},
	    \label{like_mod}\\
	    Q_{\Spec}
	    &\coloneqq
	    \sum_{i,j = 1}^{n} \left(\widehat{A}_{ij} - {P}_{ij}\right) \mathbb{I}_{\{\tau_{i} = \tau_{j}\}},
	    \label{spec_mod}\\
	    Q_{\Res}
	    &\coloneqq
	    \sum_{i,j = 1}^{n} \left({A}_{ij} - \widehat{A}_{ij}\right) \mathbb{I}_{\{\tau_{i} = \tau_{j}\}}.
	    \label{noise_mod}
	\end{align}
	\end{subequations}
\end{definition}

In \cref{def_mods3}, $Q_{\Like}$ signifies a so-called likelihood variant of modularity in which entrywise $\mathbb{E}[A_{ij} \mid \boldsymbol{\tau}] = P_{ij}$. In words, for $Q_{\Like}$ the benchmark null network corresponds to the expected graph adjacency matrix under the SBM generative model.

In \cref{def_mods3}, $Q_{\Spec}$ signifies a so-called spectral variant. This choice is motivated by the fact that for not-too-sparse SBMs, $\|\widehat{\mathbf{A}} - \mathbf{P}\|_{\max} \rightarrow 0$ in probability as $n \rightarrow \infty$ when $d$ is fixed, whereas $\|\mathbf{A} - \mathbf{P}\|_{\max}$ is bounded away from zero with probability one \cite{tang2022asymptotically}. Thus, the truncated adjacency matrix $\widehat{\mathbf{A}}$ takes the place of the raw observable data $\mathbf{A}$, while the benchmark null network $\mathbf{P}$ corresponding to the expected graph adjacency matrix is the same as in $Q_{\Like}$.

In \cref{def_mods3}, $Q_{\Res}$ signifies the residual or difference between the spectral and likelihood variants, i.e.,~$Q_{\Res} = Q_{\Spec} - Q_{\Like}$. This choice is purely data-dependent, as with the Newman--Girvan formulation. Here, the raw data matrix $\mathbf{A}$ is compared to its denoised counterpart, where these respective choices are themselves inspired by the existing literature on SBMs. This variant is further motivated by the signal-plus-noise interpretation of observed SBM adjacency matrices when written as $\mathbf{A} = \mathbf{P} + (\mathbf{A} - \mathbf{P}) = \textsf{signal} + \textsf{noise}$. Since $\widehat{\mathbf{A}}$ is a spectral matrix-valued estimator of $\mathbf{P}$, $Q_{\Res}$ juxtaposes within-community observed edges against denoised estimated signal in the edges.

\subsection{Modularity asymptotics}
\label{sec:modularity_asymptotics}

\subsubsection{Notational setup}
\label{sec:notation}

Denote the Hadamard and Kronecker products for given matrices by $\circ$ and $\otimes$, respectively. The vectorization of an $m \times n$ dimensional matrix $\mathbf{M}$ is obtained by stacking its columns to yield a vector in $\Re^{m n}$, denoted by $\operatorname{vec}(\mathbf{M})$. When $\mathbf{M}$ is symmetric, the half-vectorization is obtained by vectorizing only the lower triangular entries, including the diagonal, and is denoted by $\operatorname{vech}(\mathbf{M})$.

Denote the (skinny) spectral decomposition of $\mathbf{B}$ by $\mathbf{V}\mathbf{\Sigma}\mathbf{V}^{\top}$, where $\mathbf{V}$ is a matrix of orthonormal eigenvectors and $\mathbf{\Sigma}$ is the diagonal matrix of non-zero eigenvalues. Letting $\boldsymbol{\nu} \coloneqq \mathbf{V} |\mathbf{\Sigma}|^{1/2}$ yields the equivalent formulation $\mathbf{B} = \boldsymbol{\nu} \mathbf{I}_{p,q} \boldsymbol{\nu}^{\top}$, where $\mathbf{I}_{p,q}$ is the diagonal matrix with $p$ diagonal entries equal to $1$ and the remaining $q$ entries equal to $-1$, with $d \coloneqq \operatorname{rank}(\mathbf{B}) = p + q$. In particular, $p$ and $q$ reflect the number of positive and negative eigenvalues of $\mathbf{B}$, respectively. Define the $d \times d$ invertible matrix $\mathbf{\Delta} \coloneqq \boldsymbol{\nu}^{\top} \operatorname{diag}(\boldsymbol{\pi})\boldsymbol{\nu}$ and the idempotent matrix $\widetilde{\mathbf{\Pi}}_{\mathbf{V}}^{\perp} \coloneqq \mathbf{I} - \mathbf{V} (\mathbf{V}^{\top} \operatorname{diag}(\boldsymbol{\pi}) \mathbf{V})^{-1} \mathbf{V}^{\top} \operatorname{diag}(\boldsymbol{\pi})$. Define the vector $\widetilde{\boldsymbol{\pi}} =  \operatorname{vech}(\operatorname{diag}[\pi_{1}^{2},\dots,\pi_{K}^{2}])$ where $\operatorname{diag}([\pi_{1}^{2},\dots,\pi_{K}^{2}])$ is the diagonal matrix whose $i$-th diagonal entry is given by $\pi_{i}^{2}$. We write $\mathbf{J}$ to denote the matrix of all ones whose dimensionality is specified from context.

\subsubsection{Asymptotic distributions for modularity statistics} \label{sec:modularity_clts}

The main theoretical results in this article are presented below in the form of three theorems. Each theorem is stated in terms of population-level quantities for stochastic blockmodels, namely the number of communities $K$, the dimension $d = \operatorname{rank}(\mathbf{B})$, and the membership vector $\boldsymbol{\tau}$. The stated results continue to hold for appropriate estimates thereof, as discussed further below.

\begin{theorem}[Limiting distribution for likelihood-based modularity]
	\label{thrm:mod_like}
	For $n \ge 1$, let $\mathbf{A}^{(n)} \sim \operatorname{SBM}(\mathbf{B}, \boldsymbol{\pi})$ be a sequence of stochastic blockmodel graphs with sparsity factor $\rho_{n}$ satisfying $n\rho_{n} = \omega(\log n)$. Then, as $n \rightarrow \infty$, the likelihood variant of modularity in \cref{like_mod} satisfies
	\begin{equation}\label{like_mod_asy}
	    \rho_{n}^{-1/2}n^{-1}Q_{\Like}
	    \xrightarrow{\operatorname{d}}
	    \mathscr{N}(0, \widetilde{\boldsymbol{\pi}}^{\top} \mathbf{D}^{-1}\widetilde{\boldsymbol{\pi}}).
	\end{equation}
	The matrix $\mathbf{D}$ is specified in \cref{D-form} and depends on whether $\rho_{n} \equiv 1$ or $\rho_{n} \rightarrow 0$.
	\end{theorem}
	The matrix $\mathbf{D}$ is the $\binom{K+1}{2} \times \binom{K+1}{2}$ diagonal matrix defined entrywise as
	\begin{equation}\label{D-form}
	    {D}_{(k, l),(k, l)}
	    = 
		\begin{cases}
		    \frac{\pi_{k}\pi_{l}}{B_{kl}(1-B_{kl})(1+\mathbb{I}_{\{k=l\}})}
		    & \text{if } \rho_{n} \equiv 1,\\
		    \frac{\pi_{k}\pi_{l}}{B_{kl}(1+\mathbb{I}_{\{k=l\}})}
		    & \text{if } \rho_{n} \rightarrow 0.
		\end{cases}
\end{equation}
Above, slight abuse of notation regarding the indices of $\mathbf{D}$ is made for convenience. Namely, without loss of generality, we use the lower-triangular tuples $(1,1), (2,1), (2,2), \dots, (K,K-1), (K,K)$ to identify the entry labels $1, 2, \dots, \binom{K+1}{2}$, in an effort to clarify the correspondence with $\operatorname{vech}(\mathbf{B})$. By convention, entry $(l, k)$ equals entry $(k, l)$.

Observe that the diagonal entries of $\mathbf{D}$ are comparatively smaller in the sparse regime, $\rho_{n} \rightarrow 0$, hence the diagonal entries of $\mathbf{D}^{-1}$ are larger, resulting in a comparatively larger asymptotic variance. The interpretation here is that sparse graphs, with their fewer observable edges and hence less available data, yield in aggregate larger variability.

\begin{theorem}[Limiting distribution for spectral-based modularity]
	\label{thrm:mod_spec}
	For $n \ge 1$, let $\mathbf{A}^{(n)} \sim \operatorname{SBM}(\mathbf{B}, \boldsymbol{\pi})$ be a sequence of stochastic blockmodel graphs with sparsity factor $\rho_{n}$ satisfying $n\rho_{n} = \omega(\sqrt{n})$. Then, as $n \rightarrow \infty$, the spectral variant of modularity in \cref{spec_mod} satisfies
	\begin{equation}\label{spec_mod_asy}
	    \rho_{n}^{-1/2}n^{-1}Q_{\Spec}
	    -
	    \rho_{n}^{-1/2}\widetilde{\boldsymbol{\pi}}^{\top}\operatorname{vech}(\boldsymbol{\Theta})
	    \xrightarrow{\operatorname{d}}
	    \mathscr{N}(0, \widetilde{\boldsymbol{\pi}}^{\top} \widetilde{\mathbf{\Gamma}}\widetilde{\boldsymbol{\pi}}).
	\end{equation}
	The matrices $\mathbf{\Theta}$ and $\widetilde{\mathbf{\Gamma}}$ are specified in \cref{theta_eqs} and \cref{gamma_tilde}, respectively, with each depending on whether $\rho_{n} \equiv 1$ or $\rho_{n} \rightarrow 0$.
\end{theorem}
In the preceding theorem, the population-level bias and covariance terms are given by

\begin{align}
	\boldsymbol{\Theta}
	\coloneqq
	&\operatorname{diag}[(\mathbf{B}\circ(\mathbf{J} - \mathbf{B}))\boldsymbol{\pi}]\boldsymbol{\nu}\mathbf{\Delta}^{-1}\mathbf{I}_{p,q}\mathbf{\Delta}^{-1}\boldsymbol{\nu}^{\top} \nonumber\\
	&\quad + \boldsymbol{\nu}\mathbf{\Delta}^{-1}\mathbf{I}_{p,q}\mathbf{\Delta}^{-1}\boldsymbol{\nu}^{\top}\operatorname{diag}[(\mathbf{B}\circ(\mathbf{J} - \mathbf{B}))\boldsymbol{\pi}] \nonumber \\
	&\quad -
	\boldsymbol{\nu}\mathbf{\Delta}^{-1}\boldsymbol{\nu}^{\top} 
	\operatorname{diag}(\boldsymbol{\pi})\operatorname{diag}[(\mathbf{B}\circ(\mathbf{J} - \mathbf{B}))\boldsymbol{\pi}]\boldsymbol{\nu}\mathbf{\Delta}^{-1}\mathbf{I}_{p,q}\mathbf{\Delta}^{-1}\boldsymbol{\nu}^{\top} \nonumber \\
	&\quad -
	\boldsymbol{\nu}\mathbf{\Delta}^{-1}\mathbf{I}_{p,q}\mathbf{\Delta}^{-1}\boldsymbol{\nu}^{\top}\operatorname{diag}[(\mathbf{B}\circ(\mathbf{J} - \mathbf{B}))\boldsymbol{\pi}]\operatorname{diag}(\boldsymbol{\pi})\boldsymbol{\nu}\mathbf{\Delta}^{-1}
	\boldsymbol{\nu}^{\top}
	&& \text{~if~} \rho_{n} \equiv 1, \nonumber \\
	\boldsymbol{\Theta}
	\coloneqq
	&\operatorname{diag}[\mathbf{B} \boldsymbol{\pi}]\boldsymbol{\nu}\mathbf{\Delta}^{-1}\mathbf{I}_{p,q}\mathbf{\Delta}^{-1}\boldsymbol{\nu}^{\top} \nonumber \\
	&\quad +
	\boldsymbol{\nu}\mathbf{\Delta}^{-1}\mathbf{I}_{p,q}\mathbf{\Delta}^{-1}\boldsymbol{\nu}^{\top}\operatorname{diag}[\mathbf{B}\boldsymbol{\pi}] \nonumber \\
	&\quad - 
	\boldsymbol{\nu}\mathbf{\Delta}^{-1}\boldsymbol{\nu}^{\top} 
	\operatorname{diag}(\boldsymbol{\pi})\operatorname{diag}[\mathbf{B} \boldsymbol{\pi}]\boldsymbol{\nu}\mathbf{\Delta}^{-1}\mathbf{I}_{p,q}\mathbf{\Delta}^{-1}\boldsymbol{\nu}^{\top} \nonumber \\
	&\quad - 
	\boldsymbol{\nu}\mathbf{\Delta}^{-1}\mathbf{I}_{p,q}\mathbf{\Delta}^{-1}\boldsymbol{\nu}^{\top}\operatorname{diag}[\mathbf{B}\boldsymbol{\pi}]\operatorname{diag}(\boldsymbol{\pi})\boldsymbol{\nu}\mathbf{\Delta}^{-1}\boldsymbol{\nu}^{\top}
	&& \text{~if~} \rho_{n} \rightarrow 0, \label{theta_eqs}
\end{align}
and
\begin{equation}\label{gamma_tilde}
    \widetilde{\mathbf{\Gamma}}
    \coloneqq
    \mathcal{L}_{K}(\mathbf{I} - \widetilde{\mathbf{\Pi}}_{\mathbf{V}}^{\perp} \otimes \widetilde{\mathbf{\Pi}}_{\mathbf{V}}^{\perp})\mathcal{D}_{K} \mathbf{D}^{-1}\mathcal{D}_{K}^{\top} (\mathbf{I} - \widetilde{\mathbf{\Pi}}_{\mathbf{V}}^{\perp} \otimes \widetilde{\mathbf{\Pi}}_{\mathbf{V}}^{\perp})^{\top}\mathcal{L}_{K}^{\top} .
\end{equation}
In \cref{gamma_tilde}, $\mathcal{L}_{K}$ and $\mathcal{D}_{K}$ denote the \emph{elimination matrix} and \emph{duplication matrix}, respectively \citep{magnus1980elimination}. For any $n \times n$ symmetric matrix $\mathbf{M}$, here $\mathcal{L}_{K}$ is the unique $n(n+1)/2 \times n^{2}$ matrix satisfying $\operatorname{vech}(\mathbf{M}) = \mathcal{L}_{K}\operatorname{vec}(\mathbf{M})$. Conversely, $\mathcal{D}_{K}$ is the unique $n^{2} \times n(n+1)/2$ matrix satisfying $\operatorname{vec}(\mathbf{M}) = \mathcal{D}_{K}\operatorname{vech}(\mathbf{M})$.

\begin{theorem}[Limiting distribution for residual-based modularity]
	\label{thrm:mod_res}
	For $n \ge 1$, let $\mathbf{A}^{(n)} \sim \operatorname{SBM}(\mathbf{B}, \boldsymbol{\pi})$ be a sequence of stochastic blockmodel graphs with sparsity factor $\rho_{n}$ satisfying $n\rho_{n} = \omega(\sqrt{n})$. Further assume that $\mathbf{B}$ is strictly rank-deficient, namely $d < K$. Then, as $n \rightarrow \infty$, the residual variant of modularity in \cref{noise_mod} satisfies
	\begin{equation}\label{noise_mod_asy}
	    \rho_{n}^{-1/2}n^{-1}Q_{\Res}
	    + 
	    \rho_{n}^{-1/2}\widetilde{\boldsymbol{\pi}}^{\top}\operatorname{vech}(\boldsymbol{\Theta})
	    \xrightarrow{\operatorname{d}}
	    \mathscr{N}(0, \widetilde{\boldsymbol{\pi}}^{\top} \mathbf{\Gamma}\widetilde{\boldsymbol{\pi}}).
	\end{equation}
	The matrices $\mathbf{\Theta}$ and $\mathbf{\Gamma}$ are specified in \cref{theta_eqs} and \cref{Gamma}, respectively, with each depending on whether $\rho_{n} \equiv 1$ or $\rho_{n} \rightarrow 0$.
\end{theorem}

In \cref{thrm:mod_res}, the asymptotic covariance term $\boldsymbol{\Gamma}$ is given by
\begin{equation}
	\label{Gamma}
    \boldsymbol{\Gamma}
    \coloneqq
    \mathcal{L}_{K}(\widetilde{\mathbf{\Pi}}_{\mathbf{V}}^{\perp} \otimes \widetilde{\mathbf{\Pi}}_{\mathbf{V}}^{\perp})\mathcal{D}_{K} \mathbf{D}^{-1}\mathcal{D}_{K}^{\top} (\widetilde{\mathbf{\Pi}}_{\mathbf{V}}^{\perp} \otimes \widetilde{\mathbf{\Pi}}_{\mathbf{V}}^{\perp})^{\top}\mathcal{L}_{K}^{\top}.
\end{equation}

\cref{thrm:mod_like,thrm:mod_spec,thrm:mod_res} are stated for \cref{def_mods3} in terms of an implicit sequence of vectors $\boldsymbol{\tau}$, namely the ground truth SBM membership assignments. The theorem statements continue to hold for the estimated membership vector $\widehat{\boldsymbol{\tau}}$ derived from clustering the rows of the top $d$ eigenvectors $\widehat{\mathbf{U}} \in \Re^{n \times d}$ of $\mathbf{A}$ appearing in $\widehat{\mathbf{A}} \equiv \widehat{\mathbf{U}}\widehat{\mathbf{\Lambda}}\widehat{\mathbf{U}}^{\top}$, due to the asymptotic perfect recovery $\widehat{\boldsymbol{\tau}} = \boldsymbol{\tau}$ as discussed in \cite{tang2022asymptotically}. Furthermore, the true rank or dimension $d$ is provably consistently estimable by an eigenvalue ratio test $\widehat{d}$, since for large $n$ with high probability, $|\widehat{\lambda}_{d}|$ grows at the order $n\rho_{n}$, whereas $|\widehat{\lambda}_{d+1}|$ grows no faster than order $\sqrt{n \rho_{n}}$, i.e.,~$|\widehat{\lambda}_{d+1}| = O_{\mathbb{P}}(\sqrt{n\rho_{n}})$. For related discussion, see for example \cite{rubindelanchy2022astatistical,xie2022entrywise}.

In order to obtain the asymptotic normality results presented above, we decompose the modularity variants into entrywise estimators for $\mathbf{B}$, given below in \cref{def_block_estimate}, whose distributional properties can be precisely analyzed and aggregated. Further details are given in the supplement and rely on the following definition.

\begin{definition}[\bf Maximum likelihood estimator and spectral estimator]
	\label{def_block_estimate}
	For each $1 \le k \le K$, let $n_{k}$ denote the number of nodes in community $k$, and define the block-specific membership vector $\mathbf{s}_{k} \in \{0,1\}^{n}$ such that its $i$-th entry is $1$ if $\tau_{i} = k$ and $0$ otherwise. Let $\widehat{n}_{k}$ and $\widehat{\mathbf{s}}_{k}$ denote estimators of $n_{k}$ and $\mathbf{s}_{k}$ obtained from $\widehat{\boldsymbol{\tau}}$, derived by clustering the rows of $\widehat{\mathbf{U}}$, for example using $K$-means clustering or an expectation--maximization algorithm for fitting a mixture of $K$ Gaussians.

	Take $\widehat{\mathbf{A}}$ to have rank $d = \operatorname{rank}(\mathbf{B})$. For each block probability parameter $B_{kl}$, define the likelihood-based estimator by $\widehat{{B}}_{kl}^{(\Like)} \coloneqq \frac{1}{\widehat{n}_{k}\widehat{n}_{l}\rho_{n}}\widehat{\mathbf{s}}_{k}^{\top}\mathbf{A} \widehat{\mathbf{s}}_{l}$. Similarly, define the spectral-based estimator by $\widehat{{B}}_{kl}^{(\Spec)} \coloneqq \frac{1}{\widehat{n}_{k}\widehat{n}_{l}\rho_{n}}\widehat{\mathbf{s}}_{k}^{\top}\widehat{\mathbf{A}} \widehat{\mathbf{s}}_{l}$. The corresponding matrix-valued estimators of $\mathbf{B}$ are consequently denoted by $\widehat{\mathbf{B}}^{(\Like)}$ and $\widehat{\mathbf{B}}^{(\Spec)}$, respectively.
\end{definition}

We pause to remark that \cref{thrm:mod_like} requires the sparsity condition $n\rho_{n} = \omega(\log n)$, namely $(n \rho_{n})/(\log n) \rightarrow \infty$, whereas \cref{thrm:mod_spec,thrm:mod_res} require the stronger condition $n\rho_{n} = \omega(\sqrt{n})$. Loosely speaking, the latter requirement arises due to a spectral bias-variance trade-off stemming from the underlying aggregation of spectral estimates; see \cite{tang2022asymptotically} for more detailed discussion and an example suggesting the potential necessity of this condition.

\subsection{Hypothesis testing}
\label{sec:hypothesis_test}

The modularity variants can be used for inference by leveraging the above theorems. Supposing $\mathbf{A} \sim \operatorname{SBM}(\mathbf{B}, \boldsymbol{\pi})$ with sparsity factor $\rho_{n}$, one may wish to test hypotheses of the form
\begin{itemize}[leftmargin=*]
\centering
    \item[] $H_{0}:$ $\mathbf{B} = \mathbf{B}^{(0)}$, given the modular structure inherited by $(\boldsymbol{\tau}, \boldsymbol{\pi})$,
    \item[]
    \begin{center}
        against
    \end{center}
    \item[] $H_{1}:$ $\mathbf{B} \neq \mathbf{B}^{(0)}$, given the modular structure inherited by $(\boldsymbol{\tau}, \boldsymbol{\pi})$.
\end{itemize}

For simplicity, consider the simple alternative hypothesis with connectivity matrix $\mathbf{B}^{(1)}$. For each $i = 0, 1$, under hypothesis $H_{i}$, write $\mathbf{\Theta}^{(i)}$ corresponding to $\mathbf{\Theta}$ and $\operatorname{Var}_{\Like}^{(i)}$, $\operatorname{Var}_{\Spec}^{(i)}$ corresponding to the respective variances in \cref{thrm:mod_like,thrm:mod_spec}. Upon centering $Q_{\Like}$ and $Q_{\Spec}$ with respect to $\mathbf{B}^{(0)}$, define test statistics as follows.
\begin{enumerate}
	\label{test_statistics}
    \item $\mathrm{T}_{\Like} = \frac{\rho_{n}^{-1/2}n^{-1} Q_{\Like}}{\sqrt{\operatorname{Var}_{\Like}^{(0)}}}$;
    \item $\mathrm{T}_{\Spec} = \frac{\rho_{n}^{-1/2}n^{-1} Q_{\Spec} - \rho_{n}^{-1/2}\widetilde{\boldsymbol{\pi}}^{\top}\operatorname{vech}\boldsymbol{\Theta}^{(0)}}{\sqrt{\operatorname{Var}_{\Spec}^{(0)}}}$.
\end{enumerate}
Let $z_{\alpha}$ be the upper-$\alpha$ quantile of the standard normal distribution and $\Phi(\cdot)$ denote the standard normal cumulative distribution function. For two-sided testing with the aforementioned hypotheses, the (approximate) power functions are given as follows. For the likelihood-based variant,
\begin{equation}
	\label{Like_power}
    \mathbb{P}_{H_{1}}\left(\lvert \mathrm{T}_{\Like} \rvert > z_{\alpha /2}\right) 
    = 1 - \Phi\left(\mu_{\Like} + \sigma_{\Like} z_{\alpha/2} \right)
    + 
    \Phi\left(\mu_{\Like} - \sigma_{\Like} z_{\alpha/2} \right), 
\end{equation}
where, $\sigma_{\Like} = \sqrt{\tfrac{\operatorname{Var}_{\Like}^{(0)}}{\operatorname{Var}_{\Like}^{(1)}}}$ and $\mu_{\Like} = \tfrac{n\rho_{n}^{1/2}\sum_{k=1}^{K}\pi_{k}^{2}\left(B_{kk}^{(0)} - B_{kk}^{(1)}\right)}{\sqrt{\operatorname{Var}_{\Like}^{(1)}}}$. Similarly, for the spectral-based variant,
\begin{align}
	\label{Spec_power}
    \mathbb{P}_{H_{1}}\left(\lvert \mathrm{T}_{\Spec} \rvert > z_{\alpha /2} \right) 
    & = 1 - \Phi\left(\mu_{\Spec} + \sigma_{\Spec} z_{\alpha/2}\right) 
    + 
    \Phi\left(\mu_{\Spec} - \sigma_{\Spec} z_{\alpha/2} \right),
\end{align}
where, $\sigma_{\Spec} = \sqrt{\tfrac{\operatorname{Var}_{\Spec}^{(0)}}{\operatorname{Var}_{\Spec}^{(1)}}}$ and $\mu_{\Spec} = \tfrac{n\rho_{n}^{1/2}\sum_{k=1}^{K}\pi_{k}^{2}\left(B_{kk}^{(0)} - B_{kk}^{(1)}\right) + \rho_{n}^{-1/2}\widetilde{\boldsymbol{\pi}}^{\top}\operatorname{vech}\left(\boldsymbol{\Theta}^{(0)} - \boldsymbol{\Theta}^{(1)}\right)}{\sqrt{\operatorname{Var}_{\Spec}^{(1)}}}$.

Here, testing is dictated by the diagonal elements of the block probability matrix in conjunction with the probabilities of community assignment. For example, suppose $\mathbf{B}^{(0)}$ and $\mathbf{B}^{(1)}$ are $2 \times 2$ matrices and that they differ with respect to the $(1, 1)$ diagonal element but not with respect to the $(2, 2)$ diagonal element. Instead of testing $H_{0}: B_{11} = B_{11}^{(0)} \text{ and } B_{22} = B_{22}^{(0)}$ versus $H_{1}: B_{11} = B_{11}^{(1)} \text{ and } B_{22} = B_{22}^{(1)}$, we take into account the relative weighing effect of $\pi_{1}^{2}$ and $\pi_{2}^{2}$. Depending on whether $\pi_{1}$ is small or large, the power of the tests will range from low to high. Consequently, testing based on $\mathrm{T}_{\Like}$ and $\mathrm{T}_{\Spec}$ reflects the modular structure present in $\mathbf{A}$.

Of note, Corollary~4 in \cite{tang2022asymptotically} establishes consistent estimators for terms comprising the asymptotic bias in $Q_{\Spec}$. Additionally, one-step updates of estimators for $\mathbf{B}$ and their asymptotic efficiency properties are studied in \cite{tang2022asymptotically,xie2023efficient}.

\section{Simulations}
\label{sec:simulations}

We provide simulation examples illustrating the asymptotic theory developed for the modularity variants $\rho_{n}^{-1/2}n^{-1}Q_{\Like}$, $\rho_{n}^{-1/2}n^{-1}Q_{\Spec}$, and $\rho_{n}^{-1/2}n^{-1}Q_{\Res}$ in \cref{thrm:mod_like,thrm:mod_spec,thrm:mod_res}. For each of the following examples, we also discuss the performance of the Louvain algorithm for clustering, a choice motivated by its historical success in the community detection problem.

\subsubsection{Example~1: graphs with $K=3$, $d=3$, balanced, assortative}
\label{eg1}
\begin{figure}[ht]
	\includegraphics[width=9cm, height = 2.5cm]{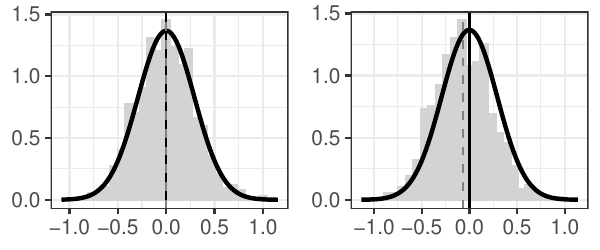}
	\centering
	\caption{Dense networks in \cref{eg1} with $n = 300$ nodes. Left plot shows $\rho_{n}^{-1/2}n^{-1}Q_{\Like}$, and right plot shows $\rho_{n}^{-1/2}n^{-1}Q_{\Spec}$. Dashed vertical line shows bias in simulation. Solid vertical line shows population bias. Solid curve shows population density fit.}
	\label{eg1_dense_300}
\end{figure}

First, we consider a three-block SBM with block-probability matrix having affinity structure. We simulate $1000$ independent replicates for each value $n$ in the setting
\begin{equation}
    \mathbf{B}
    =
    \begin{bmatrix}
		0.85 & 0.50 & 0.25\\
		0.50 & 0.85 & 0.50\\
		0.25 & 0.50 & 0.85
	\end{bmatrix}, 
	\quad 
	\boldsymbol{\pi}
	=
	\begin{bmatrix}
		1/3\\
		1/3\\
		1/3
	\end{bmatrix},
	\quad
	n \in \left\{300,600,1800,6000\right\}.
\end{equation}
We present the dense case, $\rho_{n} \equiv 1$, as well as a sparse regime, $\rho_{n} = n^{-1/4}$. In both the dense and sparse regimes, the theoretical asymptotic bias is zero. For the dense regime, relatively small networks already exhibit approximate normality in simulations. When $n = 300$ (see \cref{eg1_dense_300}), expected block sizes are $100$, and the theoretical variances and bias are already well illustrated by the $1000$ replicates; bias and variance estimates for different network sizes are shown in \cref{eg1_table_dense}. The Louvain algorithm, which uses $Q_{\Ng}/2m$ as the modularity function, accurately recovers the communities with adjusted Rand index (ARI) \citep{hubert1985comparing} near one. The latter observation is unsurprising given the assortative (homophilic) network structure present here.

\begin{table}[ht]
    \begin{center}
        \begin{tabular}{| c | c | c | c | c | c | c | c | c | }
        \hline
        $n$
            & \multicolumn{4}{c|}{$\rho_{n}^{-1/2}n^{-1}Q_{\Like}$} & \multicolumn{4}{c|}{$\rho_{n}^{-1/2}n^{-1}Q_{\Spec}$}
            \\
            \cline{2-9}
               &   \multicolumn{2}{c|}{Bias}  &   \multicolumn{2}{c|}{Variance}
               &   \multicolumn{2}{c|}{Bias}  &   \multicolumn{2}{c|}{Variance}
            \\
            \cline{2-9}
              &   Theory  &   Simulation &   Theory  &   Simulation &   Theory  &   Simulation &  Theory  &   Simulation
            \\
            \hline
            $300$ &  $0$  & $-0.0038$  & $0.085$  &  $0.0866$ & $0$ & $-0.0690$  & $0.085$  & $0.0887$
            \\
            \hline
            $600$ &  $0$  &  $0.0122$ & $0.085$   & $0.0865$ & $0$ & $-0.0216$ & $0.085$ &  $0.0867$ 
            \\
            \hline
            $1800$ &  $0$  &  $0.0052$ &  $0.085$  & $0.0845$ & $0$ & $-0.0061$ & $0.085$ &  $0.0845$ 
            \\
            \hline
            $6000$ & $0$  &  $0.0000$ &  $0.085$  & $0.0880$ & $0$ & $-0.0034$ & $0.085$ &  $0.0880$ 
            \\
            \hline
        \end{tabular}
    \caption{Bias and variance for dense case in \cref{eg1}. Four decimal places are shown.}
    \label{eg1_table_dense}
    \end{center}
\end{table}

\begin{figure}[ht]
	\includegraphics[width=10cm]{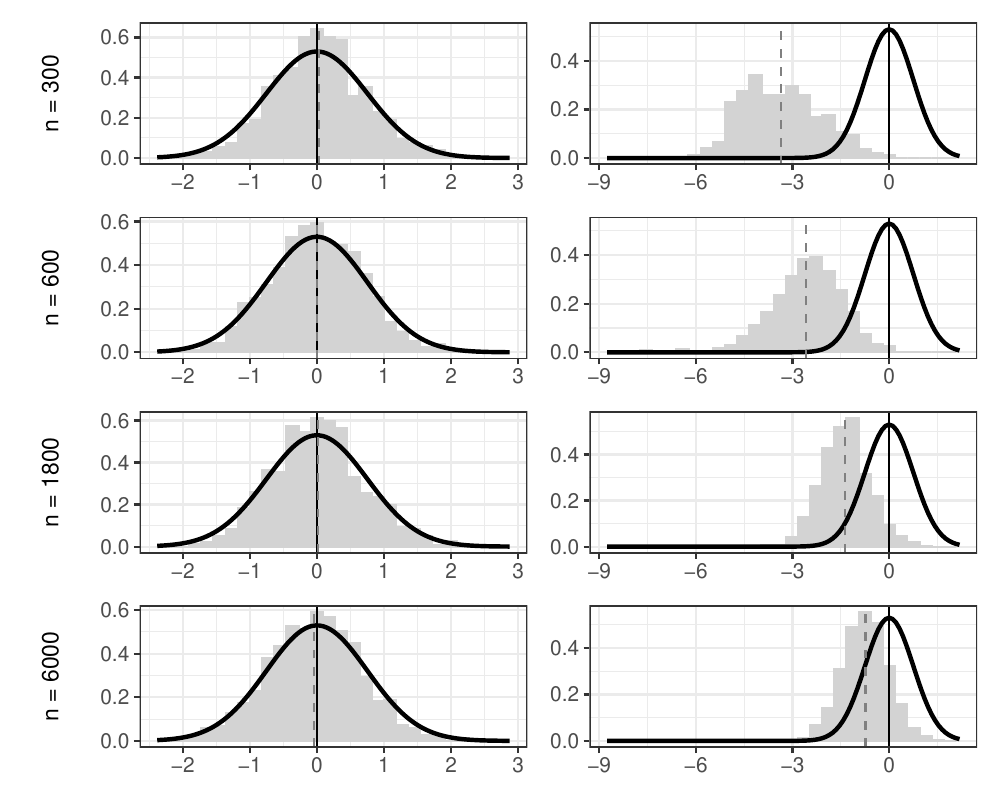}
	\centering
	\caption{Sparse networks in \cref{eg1} for $n \in \{300,600,1800,6000\}$ nodes. Left panel shows $\rho_{n}^{-1/2}n^{-1}Q_{\Like}$, and right panel shows $\rho_{n}^{-1/2}n^{-1}Q_{\Spec}$. Dashed vertical line shows bias in simulation. Solid vertical line shows population bias. Solid curve shows population density fit.}
	\label{eg1_sparse}
\end{figure}

Results for the sparse regime are illustrated in \cref{eg1_sparse}. For the spectral variant, $n = 6000$ still corresponds to a small effective node sample size, resulting in the persistence of noticeable discrepancy between simulation and theory. For simulations with larger values of $n$, say $n \in \{9000,12000\}$, the discrepancy continues to decrease (not shown), as anticipated by theory. The Louvain clustering method performs moderately well already when the network size is $300$, somewhat overestimating the number of clusters, with ARI values less than $0.8$ and on average near $0.5$. As the network size increases, the clustering and estimated number of clusters approach the ground truth.

\subsubsection{Example~2: graphs with $K=2$, $d=2$, balanced, disassortative}
\label{eg2}
\begin{figure}[ht]
	\includegraphics[width=10cm]{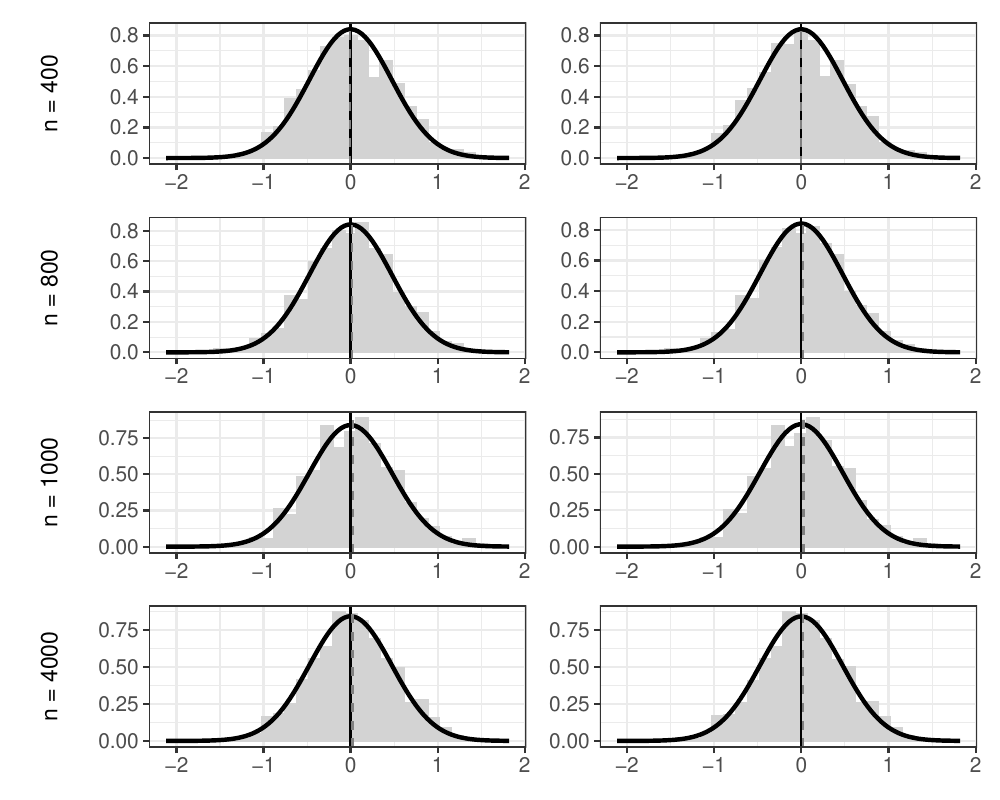}
	\centering
	\caption{Dense networks in \cref{eg2} with $n \in \{400,800,1000,4000\}$ nodes. Left panel shows $\rho_{n}^{-1/2}n^{-1}Q_{\Like}$, and right panel shows $\rho_{n}^{-1/2}n^{-1}Q_{\Spec}$. Dashed vertical line shows bias in simulation. Solid vertical line shows population bias. Solid curve shows population density fit.}
	\label{eg2_dense}
\end{figure}

Consider the dense two-block SBM setting with
\begin{equation}
    \mathbf{B}
    =
    \begin{bmatrix}
        0.30 & 0.75\\
        0.75 & 0.40
    \end{bmatrix},
    \quad
    \boldsymbol{\pi} 
    =
    \begin{bmatrix}
        1/2\\
        1/2
	\end{bmatrix},
	\quad
    n \in \left\{400,800,1000,4000\right\}.
\end{equation}
Graphs generated from this model exhibit so-called disassortative structure, meaning that nodes in different blocks are more likely to be connected than nodes within the same block. For each choice of network size $n$, simulations are conducted over $1000$ independent replications. For $n = 400$, the theoretical values of the parameters are already well estimated by the simulated data per \cref{eg2_dense}. The Louvain algorithm shows poor performance in detecting clusters which is anticipated and due to the disassortative network connectivity structure. Specifically, the number of detected clusters ranges from five to eleven, with near-zero ARI values.

\subsubsection{Example~3: graphs with $K=2$, $d=1$, unbalanced, core-periphery}
\label{eg3}
\begin{figure}[ht]
	\includegraphics[width=10cm, height = 8cm]{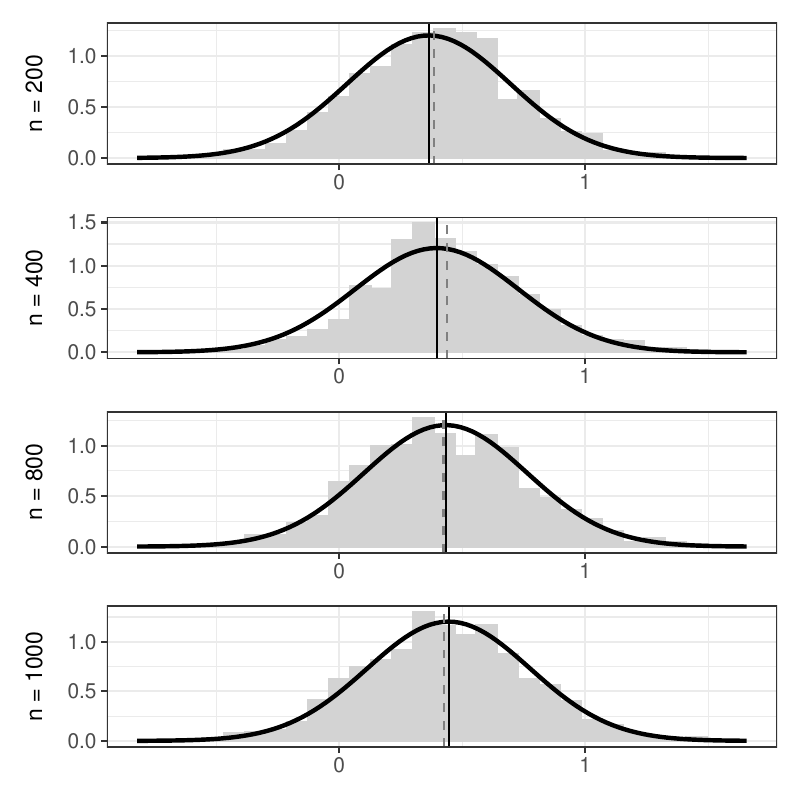}
	\centering
	\caption{Sparse networks and residual-based modularity in \cref{eg3}. Dashed vertical line shows bias in simulation. Solid vertical line shows population bias. Solid curve shows population density fit.}
	\label{eg3_resid}
\end{figure}

Consider the two-block SBM where $\mathbf{B}$ is rank one of the form $\left[\begin{smallmatrix} p\\ q \end{smallmatrix}\right] \times 
\left[\begin{smallmatrix} p & q \end{smallmatrix} \right]$. Specifically,
\begin{equation}
    p=3/4,\; q=1/4,
    \quad
    \mathbf{B}
    =
    \begin{bmatrix}
        0.5625 & 0.1875\\
        0.1875 & 0.0625
	\end{bmatrix},
    \quad
    \boldsymbol{\pi}
    =
    \begin{bmatrix}
        1/4\\
        3/4
    \end{bmatrix},
    \quad
    n \in \{200,400,800,1000\},
\end{equation}
and set $\rho_{n} = n^{-1/4}$. In this example, $\mathbf{B}$ exhibits core-periphery structure. \cref{eg3_resid} shows that even in this sparse setup with relatively small network sizes the simulation performance supports the asymptotic theory. Here simulations are presented for $1000$ independent trials, as above. Due to the underlying core-periphery network structure, the Louvain algorithm fails to detect the clusters and always vastly over-estimates the true number of clusters. 

\subsubsection{Contour plots and parameter space}
This section briefly illustrates how the asymptotic bias and variance of modularity values change as a function of different SBM parameter choices. Two different examples of two-block SBMs are presented, both with $\rho_{n} \equiv 1$.

\begin{figure}[ht]
	\includegraphics[width=10cm, height=10cm]{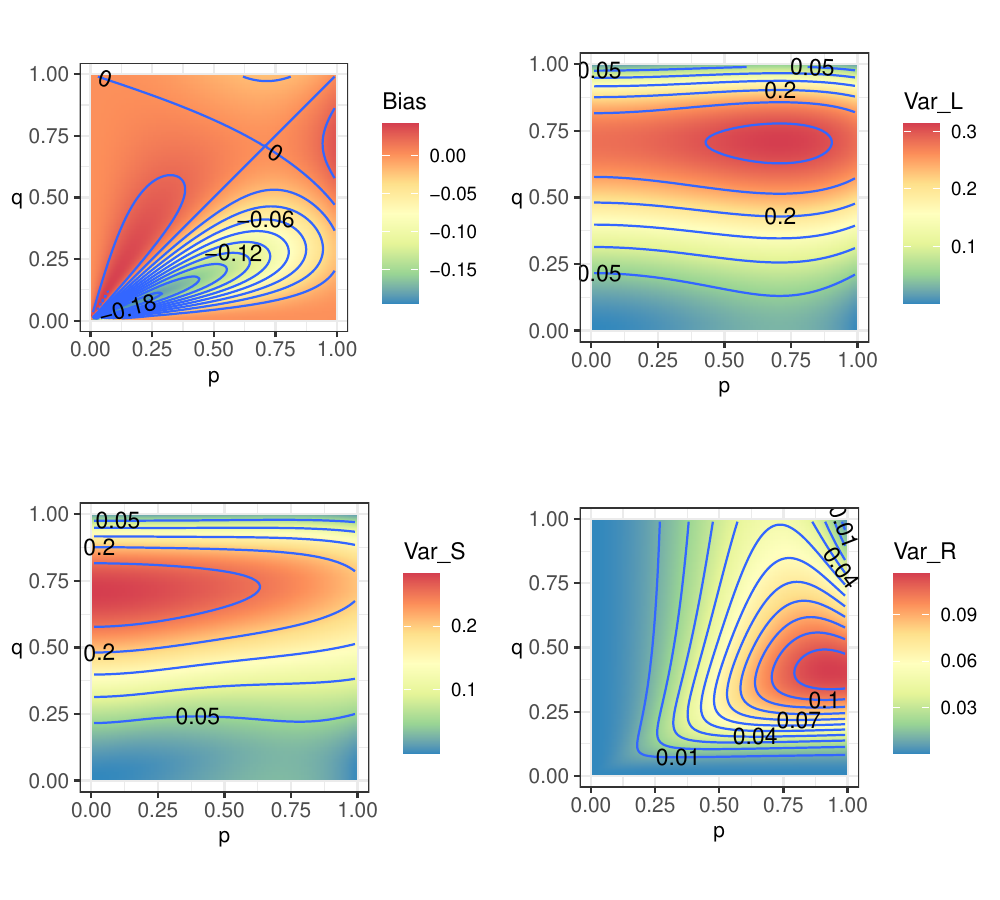}
	\centering
	\caption{Asymptotic bias and variance plotted as functions of $(p,q)$ where $\mathbf{B} = \left[\begin{smallmatrix} p^{2} & pq\\ pq & q^{2} \end{smallmatrix}\right]$ and $\boldsymbol{\pi} = [1/4, 3/4]^{\top}$.}
	\label{Eg_A_contour}
\end{figure}

\begin{figure}[ht]
	\includegraphics[width=10cm, height=10cm]{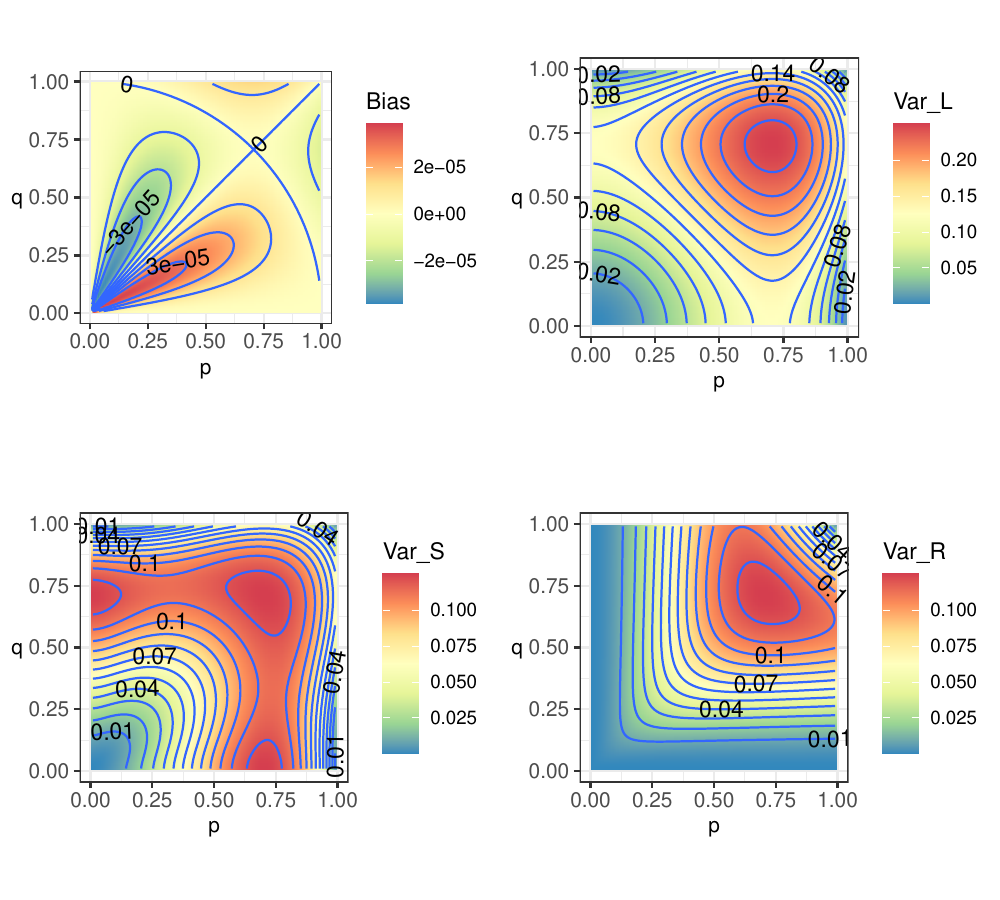}
	\centering
	\caption{Asymptotic bias and variance plotted as functions of $(p,q)$ where $\mathbf{B} = \left[\begin{smallmatrix} p^{2} & pq\\ pq & q^{2} \end{smallmatrix}\right]$ and $\boldsymbol{\pi} = [0.5001, 0.4999]^{\top}$.}
	\label{Eg_A_contour_equal}
\end{figure}

In the first example, \cref{Eg_A_contour}, $\mathbf{B}$ is of the form $\left[\begin{smallmatrix} p\\ q \end{smallmatrix}\right] \times \left[\begin{smallmatrix} p & q \end{smallmatrix} \right]$. Setting $\boldsymbol{\pi} = [1/4, 3/4]^{\top}$, we plot the asymptotic bias, $\widetilde{\boldsymbol{\pi}}^{\top}\operatorname{vech}(\boldsymbol{\Theta})$, and the asymptotic variances from \cref{thrm:mod_like,thrm:mod_spec,thrm:mod_res} as functions of $(p,q)$. In contrast, when setting $\boldsymbol{\pi} = [1/2, 1/2]^{\top}$, the block-specific biases cancel out to yield an overall bias of zero. This happens because the bias corresponding to the first block, with $p^{2}$, is the negative of that for the second block, with $q^{2}$. This property is exhibited along with the variances in \cref{Eg_A_contour_equal} where we choose $\boldsymbol{\pi} = [0.5001, 0.4999]^{\top}$. For $Q_{\Like}$, the variance is maximal when both $p$ and $q$ are near $0.75$ as exhibited in \cref{Eg_A_contour}. For moderately large values of $q$, the variance may still be large even if $p$ is small which is due to our choice of $\boldsymbol{\pi}$. In contrast, for $Q_{\Spec}$, the variance is larger if $q$ is near $0.65$ with smaller values of $p$. The variance corresponding to $Q_{\Res}$ increases as $p$ increases when $q$ is near $0.3$. As such, loosely speaking, it negotiates the difference between the variances of $Q_{\Like}$ and $Q_{\Spec}$. Similar observations follow from \cref{Eg_A_contour_equal}. Importantly, the balanced block structure in this scenario induces symmetry in the plots for variances of $Q_{\Like}$, $Q_{\Spec}$, and $Q_{\Res}$.

\begin{figure}[ht]
	\includegraphics[width=10cm, height=10cm]{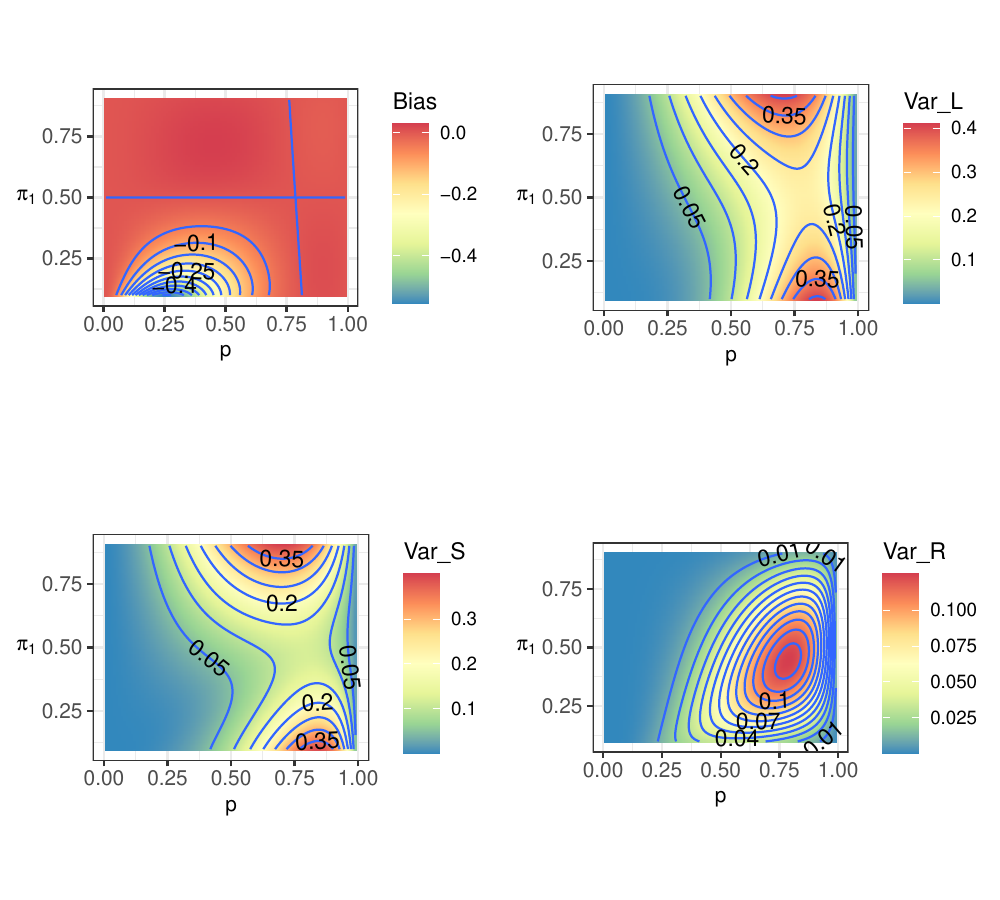}
	\centering
	\caption{Asymptotic bias and variance plotted as functions of $(p,\pi_{1})$ where $\mathbf{B} = \left[\begin{smallmatrix}
	  p^{2} & p^{3}\\
	  p^{3} & p^{4}
	\end{smallmatrix}\right]$ and $\boldsymbol{\pi} = [\pi_{1}, 1-\pi_{1}]^{\top}$.}
	\label{Eg_D_contour}
\end{figure}

In the second example, \cref{Eg_D_contour}, $\mathbf{B}$ is of the form $\left[\begin{smallmatrix} p\\ p^{2} \end{smallmatrix}\right] \times \left[\begin{smallmatrix} p & p^{2} \end{smallmatrix} \right]$ and $\boldsymbol{\pi} = [\pi_{1}, 1-\pi_{1}]^{\top}$. In the same manner we study the asymptotic bias and variances as functions of $(p, \pi_{1})$. The bias lies approximately in the range from $-0.554$ to $0.026$, with most values being larger than $0.05$. Irrespective of $p$, the bias is zero when $\pi_{1} = 1/2$ which can be analytically verified in broader generality. We further see that for both $Q_{\Like}$ and $Q_{\Spec}$, variance is large when $p$ is large and $\pi_{1}$ is close to either zero or one. Their contours exhibit overall similar landscapes. The variance profile of $Q_{\Res}$ behaves noticeably differently, exhibiting a single mode near $p=0.8$, $\pi_{1}=0.4$.

\section{Real data analysis}
\label{sec:real_data}

This section applies all three modularity variants to the study of network neuroscience. Each dataset under investigation consists of collections of human brain networks on the same analysis-specific node set but with different edge connectivity properties.

\subsection{Data preliminaries}

Modularity is commonly used to explain community structure manifest in anatomical and functional brain connectivity. For example, it has been employed to compare healthy subjects to patients diagnosed with schizophrenia or Alzheimer's disease \citep{van2014brain,contreras2019resting,zhang2021modularity}. Modularity quantifies the strength of discernible community structure, while the network property known as small-worldness is defined by high local clustering and short global links; thus, a decrease in modularity might also indicate a disruption of small-worldness. \cite{esfahlani2021modularity} discusses various types of possible null models for different definitions of brain networks and consequent adjustments to the modularity function. \cite{alexanderbloch2010disrupted} reports reductions in local connectivity and small-worldness in childhood-onset schizophrenia patients. These findings conform with the findings in \cite{bullmore2009complex} that, compared to healthy individuals, small-world properties seem to degrade in individuals with schizophrenia. \cite{sporns2016modular} reviews numerous findings on brain networks where functional correlation networks are treated in different ways and modularity optimization is applied for community detection.

In the study of brain networks, nodes are commonly referred to as regions of interest (ROIs). Aside from the specification of biologically or structurally determined parcellations of brain networks, community structure among ROIs is typically unknown \emph{a priori} and must therefore be inferred. Further, networks derived from brain imaging data typically contain edges with weights and signs (i.e.,~non-binary, positive or negative). A common technique used to remove seemingly `noisy' or `false positive' edges is to threshold and binarize networks. Among existing approaches, \cite{bordier2017graph} considers a percolation-based approach for determining theshold selection. Another technique for obtaining binary graphs is the $k$-nearest neighbor construction computed on the basis of node feature distances \citep{luxburg2007tutorial}. We, too, must perform binarization in this section as a data preprocessing step.

\subsubsection{COBRE data}
\label{sec:cobre_data}

We consider the COBRE dataset of brain networks described in \cite{aine2017multimodal} which is freely available in the public database \href{http://fcon_1000.projects.nitrc.org/indi/retro/cobre.html}{http://fcon\_1000.projects.nitrc.org/indi/retro/cobre.html}. It consists of networks with $263$ regions of interest (ROIs) formed from functional magnetic resonance imaging (fMRI) of brains from (i) $70$ healthy control subjects (HC) and (ii) $54$ patients with either schizophrenia or schizoaffective disorder (SZ). We make use of the parcellation given by \cite{power2011functional} which divides the ROIs into 14 functional brain systems. Specifically, we consider the preprocessed data used in \cite{arroyo2019network} available at \href{https://github.com/jesusdaniel/graphclass}{https://github.com/jesusdaniel/graphclass}. By construction, the data are represented by correlation matrices which, after removing the main diagonal, can be viewed as adjacency matrices of loop-free edge-weighted graphs. We entrywise binarize the absolute correlations at the threshold value $0.3$, a compromise between removing potentially noisily-observed edges while still preserving the connectivity of all networks in the sample. In particular, this choice derives from the observation that $0.31$ and $0.34$ are the largest threshold values preserving connectivity among all healthy controls and patients, respectively. Further, the threshold choice $0.3$ corresponds to per-subject absolute correlation percentile values between $75\%$ and $86\%$.

To establish a baseline, we begin with a more traditional approach to modularity-based analysis using the Newman--Girvan specification, $Q_{\Ng}$. For each subject, we run the \textsf{leiden()} modularity maximization routine in \textsf{R} for one hundred different seed-based initializations (replicates). As with the Louvain algorithm, the Leiden algorithm is but one of many popular approaches for modularity-based analysis. Doing so yields one hundred partitions of the node set (ROIs) for each subject. The replicate-replicate pairwise ARI values vary between $0.3$ and $0.9$, indicating sensitivity to initialization and imperfect agreement across replicates, yet they often exceed $0.5$ and thus are often not too dissimilar from one another. The corresponding modularity values themselves exhibit heterogeneity per subject, yet almost all subject-specific distributions of modularity values exhibit large negative skew which is consistent with the aim of the underlying algorithm to maximize modularity. Finally, for each subject, we retain their largest Newman--Girvan modularity value and the corresponding partition yielding a node clustering membership vector.

We find that although the mean, median, and standard deviation of the HC modularity values nominally exceed those of the SZ modularity values, their distributions are not significantly different based on a two-sided two-sample Kolmogorov--Smirnov test ($p\text{-value} > 0.5$). Marginally, for both HC and SZ, the distribution of modularity values fails to reject the null hypothesis of normality based on the Shapiro--Wilk test ($p\text{-value} > 0.25$). Further, two-sided $t$-tests under either the equal or unequal variance specification provide weak evidence that the true difference in means equals zero ($p\text{-value} > 0.6$). These findings are not entirely surprising, given the amount of preprocessing needed for this analysis and the relatively small sample sizes for both controls and patients. Further, the present design and testing setup may also be underpowered.

Next, we examine how the above analysis is impacted by modifying the choice of null network. Our first observation is that $Q_{\Ng}$, wherein $P_{ij} = k_{i}k_{j}/(2m)$, gives a rank-one choice of null network based on the graph degree profile. On the other hand, the rank-one approximation $\widehat{\mathbf{A}} = \widehat{\lambda}_{1}\widehat{\mathbf{u}}_{1}\widehat{\mathbf{u}}_{1}^{\top}$ similarly reflects the degree profile or row sums of $\mathbf{A}$. Consequently, up to suitable scaling, for this choice of $\widehat{\mathbf{A}}$ it is anticipated that $Q_{\Ng} \approx Q_{\Res}$. Indeed, we observe this behavior for both controls and patients, namely that the rank-one residual-based modularity values do not systematically differ from $Q_{\Ng}$. Further, each marginal distribution of modularity values is not significantly non-Gaussian; this is anticipated on the basis of our theory provided the underlying adjacency matrices are at least approximately block-structured. By choosing to increase the low-rank approximation of $\mathbf{A}$ to dimension two and three, namely by considering $Q_{\Res}$ with null networks $\sum_{i=1}^{2}\widehat{\lambda}_{i}\widehat{\mathbf{u}}_{i}\widehat{\mathbf{u}}_{i}^{\top}$ and $\sum_{i=1}^{3}\widehat{\lambda}_{i}\widehat{\mathbf{u}}_{i}\widehat{\mathbf{u}}_{i}^{\top}$, respectively, the modularity distributions now appreciably differ from $Q_{\Ng}$ yet remain approximately Gaussian.

A general challenge facing the use of spectral methods is that between-subject heterogeneity can lead to different estimates for dimensionality, the number of blocks, and block sizes, yielding downstream incompatibilities and complicating comparisons. Indeed, our initial attempts at dimension selection via eigenvalue ratio tests or the method of profile likelihood \citep{zhu2006automatic} yielded substantial subject-subject differences among controls and patients. Instead, for each network adjacency matrix, we apply the recent method for estimating graph dimension based on cross-validated eigenvalues in \cite{chen2021estimatingARXIV} as implemented in the \textsf{R} library \textsf{gdim} and function \textsf{eigcv()}. This flexible spectral method is designed for approximately low rank and block-structured data yet is not restricted to stochastic blockmodels or even parametric models; it does not cluster the nodes (ROIs) but instead returns the number of informative dimensions (eigenvectors) for input (adjacency) matrices. For both controls and patients, the subject-specific estimated dimension values range from seven to sixteen and concentrate around the values eleven and twelve. This finding is striking and encouraging based on the fact that the underlying Power parcellation (atlas-based partition) divides the ROIs into thirteen known systems and one `\emph{Uncertain}' system but where several systems such as \emph{Cerebellar} and \emph{Memory retrieval} have far fewer nodes than others such as the \emph{Default mode} or \emph{Visual} systems (see the supplement). In the present setting, the \emph{a posteriori} mean, median, and mode estimated dimensionality values for both HC and SZ are closely supported by the \emph{a priori} known Power parcellation partitioning of the $263$ brain regions.

On the basis of both the atlas-based parcellation and spectral-based estimated dimensionality of the networks, we next investigate properties of $Q_{\Like}$, $Q_{\Spec}$, and $Q_{\Res}$ for the COBRE data, reported in  \cref{table:data_COBRE}. We begin by computing the per-population average adjacency matrices $\overline{\mathbf{A}}_{\operatorname{HC}}$ and $\overline{\mathbf{A}}_{\operatorname{SZ}}$ along with the corresponding estimated connectivity matrices $\widehat{\mathbf{B}}_{\operatorname{HC}}^{14 \times 14}$ and $\widehat{\mathbf{B}}_{\operatorname{SZ}}^{14 \times 14}$ using $\boldsymbol{\tau}_{\operatorname{POWER}}$ and supposing that $\rho_{n} \equiv 1$ (reported behavior is similar for smaller values of $\rho_{n}$). We treat these averages as proxies for the corresponding population level quantities $\mathbf{P}_{\operatorname{CLASS}}$ and $\mathbf{B}_{\operatorname{CLASS}}$ for $\operatorname{CLASS} \in \{\operatorname{HC}, \operatorname{SZ}\}$. Similarly, $\boldsymbol{\tau}_{\operatorname{POWER}}$ together with $n$ leads us to write $\boldsymbol{\pi}_{\operatorname{HC}} = \boldsymbol{\pi}_{\operatorname{SZ}}$. On the basis of this block-wise projection with $\widehat{K}=\widehat{d}=14$, we plug in the estimated connectivity values into our asymptotic formulas to get estimated bias and variance values. Since several of the fourteen communities (systems) have only a few nodes, we further consider a coarser partition of the ROIs by grouping together brain systems (full details provided in the supplement). Doing so yields $\widehat{K}=5$ blocks, each with more nodes hence larger effective sample sizes. The matrices $\widehat{\mathbf{B}}_{\operatorname{HC}}^{5 \times 5}$ and $\widehat{\mathbf{B}}_{\operatorname{SZ}}^{5 \times 5}$ each have two eigenvalues that are at least one order of magnitude smaller than the largest three eigenvalues, thereby motivating the investigation of rank $\widehat{d} \in \{3, 4, 5\}$ truncations of each connectivity matrix. Overall, \cref{table:data_COBRE} shows similarities for both populations across different settings, with minor reductions in estimated variability among healthy controls compared to patients.

\begin{table}[t]
\begin{center}
	\begin{tabular}{||c c c c c c c||} 
		\hline
		Type & $\widehat{K}$ & $\widehat{d}$ & $\widehat{\text{Bias}}$ & $\widehat{\sigma}_{\Like}^{2}$ & $\widehat{\sigma}_{\Spec}^{2}$ & $\widehat{\sigma}_{\Res}^{2}$ \\ [0.5ex] 
		\hline\hline
		HC & 14 & 14 & 0 & 0.049 & 0.049 & NA \\ 
		SZ & 14 & 14 & 0 & 0.049 & 0.049 & NA \\ 
		\hline
		HC & 5 & 5 & 0 & 0.105 & 0.105 & NA \\
		SZ & 5 & 5 & 0 & 0.107 & 0.107 & NA \\
		\hline
		HC & 5 & 4 & 0.025 & 0.100 & 0.092 & 0.005 \\
		SZ & 5 & 4 & 0.041 & 0.101 & 0.092 & 0.006 \\
		\hline
		HC & 5 & 3 & 0.019 & 0.097 & 0.084 & 0.011 \\
		SZ & 5 & 3 & 0.018 & 0.098 & 0.084 & 0.012 \\
		\hline
	\end{tabular}
\end{center}
 \caption{Modularity parameter estimates for COBRE data using the Power parcellation. Reported values are rounded.}
\label{table:data_COBRE}
\end{table}

\subsubsection{UK Biobank data}
\label{sec:UKB_data}

The UK Biobank is ``a large-scale biomedical database and research resource containing genetic, lifestyle and health information from half a million UK [United Kingdom] participants'' with additional information available at \href{http://ukbiobank.ac.uk}{http://ukbiobank.ac.uk}. We consider resting state fMRI scans for $450$ healthy controls (HC) and $469$ patients (PT) affected with psychosis, including but not limited to being diagnosed with schizophrenia. We investigate subject-specific networks consisting of $377$ ROIs classified into $23$ biologically-justified modules \citep{glasser2016multi}. The second author's lab group suggests treating these modules as `experimental truths' or `literature truths', rather than as `ground truths,' which motivates the analysis herein to avoid directly comparing data-derived partitions to these modules. In the interest of parsimony, we treat $23$ as a contextually-informed upper bound on the number of clusters for all networks.

Motivated by \cite{lei2022graph}, we first Fisher-transform the entries of the functional correlation networks and subsequently obtain binary $50$-nearest neighbor networks. This approach produces directed graphs which are subsequently symmetrized such that two nodes are connected when at least one of them is a nearest neighbor of the other. Across all healthy controls, node degrees take values in the interval $[86, 161]$, whereas the corresponding interval for patients is $[90, 155]$.

For the sake of benchmarking, we first examine the properties of $Q_{\Ng}$ using the \textsf{leiden()} modularity maximization routine across twenty different initializations per subject, keeping the largest modularity value and corresponding partition (node clustering) for each. We find that the mean and median modularity values are nominally larger for the healthy control group than for the patient group; further, the standard deviation of modularity is nominally smaller for the healthy control group. However, the modularity distributions are not significantly different based on a two-sided two-sample Kolmogorov--Smirnov test ($p\text{-value} > 0.7$); neither distribution fails to reject normality based on the Shapiro--Wilk test ($p\text{-values} > 0.05$), and the distributions display an insignificant difference in means based on a two-sample $t$-test ($p\text{-value} > 0.8$). These preliminary investigations provide weak evidence of differences in modularity values between groups.

\begin{figure}[t]
	\includegraphics[width=12cm, height=7cm]{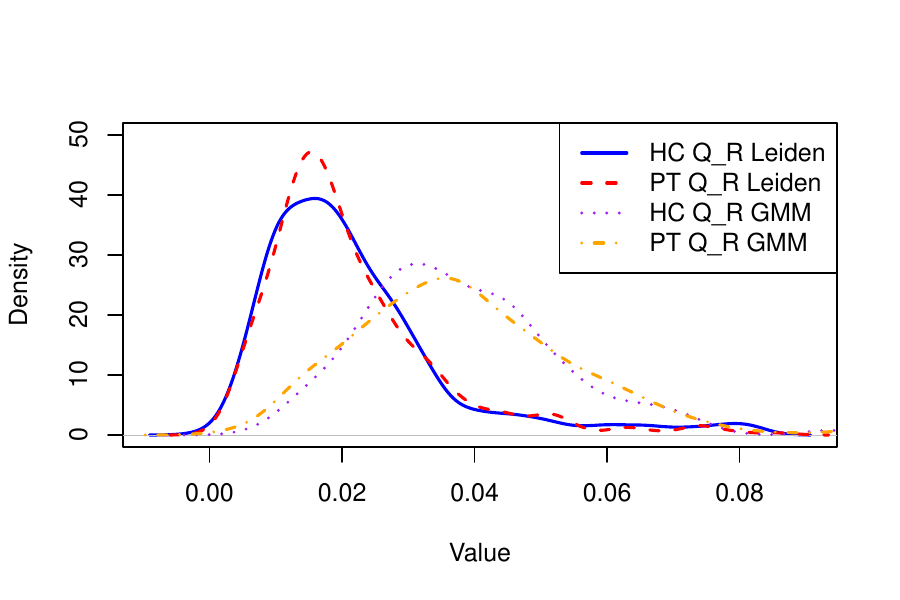}
	\centering
	\caption{Estimated residual-based modularity densities for the UK Biobank for different choices of node clustering.}
	\label{fig:UKB_density}
\end{figure}

Here, the \textsf{leiden()} clustering algorithm returns $\widehat{K} \in \{5,6\}$ across both healthy controls and patients, each with mode equal to $5$. Using the \textsf{R} library \textsf{gdim} and function \textsf{eigcv()} yields estimated graph dimension values $\{11,\dots,15\}$ for controls and $\{11,\dots,16\}$ for patients, each with mode equal to $13$ and second-most-frequent value equal to $14$. Interestingly, here $\widehat{K} < \widehat{d}$ which lies outside the set of possibilities for stochastic blockmodel graphs. Nevertheless, we can still ask about the behavior of $Q_{\Res}$ for $\widehat{d}=5$. In particular, we consider two choices for $\widehat{\boldsymbol{\tau}}$, the first based on clustering with \textsf{leiden()}, and the second based on Gaussian mixture model clustering with five components applied to the leading $\widehat{K}$ eigenvectors of each adjacency matrix, using the \textsf{R} functionality \textsf{Mclust()}. \cref{fig:UKB_density} displays the results, showing that unlike above, here the distributions of modularity values significantly deviate from normality, yet the distributional differences between groups are not appreciably different for either clustering method.

Our analysis suggests that the control sample is not significantly different from the patient sample in terms of several modularity-based characteristics. This conclusion is somewhat expected and agrees with precedent, given that (i) the data are from resting state fMRI rather than task-based fMRI and that (ii) the patient population is aggregated across different, varied psychosis diagnoses, which include the following classifications: schizophrenia, schizoaffective disorder, recurrent depression, manic episodes, bipolar disorder, and unusual psychotic experience. Our findings are in terms of multiple approaches to modularity-based analysis, rather than just a single modularity criterion, enabling nuance and inquiry from different points of view.

\section{Discussion}
\label{sec:discussion}

The findings in this paper can be readily applied together with existing results for stochastic blockmodel graphs and their known statistical properties. For example, \cite{lei2016goodnessoffit} establishes a goodness of fit criterion for determining the number of communities in SBM graphs. \cite{tang2022asymptotically} provides a comparison of likelihood and spectral approaches for SBM parameter estimation. \cite{zhang2022randomized} provides guarantees for large-scale, randomized SBM analysis. Further, SBM model selection can in principle be carried out on observed data via the network cross-validation approach in \cite{chen2018network} or the edge cross-validation approach in \cite{tianxi2020network}, though these methods do not reach a consensus when applied to our data. Of note, recent years have witnessed significant advances regarding the problem of estimating the number of communities $K$ in network models \citep{jin2023optimal,han2023universal,chen2021estimatingARXIV,hwang2023estimation}. Moving forward, these developments can conceivably be incorporated in the ongoing study of modularity and its variants, thereby avoiding the problematic yet still oft-posited assumption that $K$ is known.

In practice, the choice of modularity variant \emph{may} or arguably necessarily \emph{should} be application-dependent, a sentiment shared in \cite{bazzi2016community}. Here we strive for illustrations, rather than definitive analysis of any one particular dataset, hence we consider multiple modularity variants in tandem. Even when faced with relatively small networks, asymptotics may still be useful as we have attempted to demonstrate.

Of course, blockmodels are not themselves the end goal when investigating complex networks. Nevertheless, they offer a tractable starting point for investigating and developing statistical foundations. Encouragingly, (at least) approximate block structure is often observed empirically or biologically plausible and can be productively leveraged to unify theory, methods, and practice \citep{priebe2019twotruths}.

\section*{Supplementary materials}
\begin{description}
	\item[Appendix:] Document containing proofs of results in the article, additional simulation examples, and further discussion of the data analysis.
	\item[R code and data:] Code to reproduce the simulations and real data analysis in the article, subject to data sharing permission.
\end{description}

\clearpage
\bibliographystyle{plainnat}
\bibliography{refs}

\begin{thebibliography}{84}
\providecommand{\natexlab}[1]{#1}
\providecommand{\url}[1]{\texttt{#1}}
\expandafter\ifx\csname urlstyle\endcsname\relax
  \providecommand{\doi}[1]{doi: #1}\else
  \providecommand{\doi}{doi: \begingroup \urlstyle{rm}\Url}\fi

\bibitem[Abbe(2018)]{abbe2018community}
Emmanuel Abbe.
\newblock Community detection and stochastic block models: recent developments.
\newblock \emph{Journal of Machine Learning Research}, 18\penalty0
  (177):\penalty0 1--86, 2018.
\newblock URL \url{http://jmlr.org/papers/v18/16-480.html}.

\bibitem[Aicher et~al.(2014)Aicher, Jacobs, and Clauset]{aicher2014learning}
Chrisopher Aicher, Abigail~Z. Jacobs, and Aaron Clauset.
\newblock Learning latent block structure in weighted networks.
\newblock \emph{Journal of Complex Networks}, 3\penalty0 (2):\penalty0
  221--248, Jun 2014.
\newblock \doi{10.1093/comnet/cnu026}.
\newblock URL \url{https://doi.org/10.1093%2Fcomnet%2Fcnu026}.

\bibitem[Aicher et~al.(2013)Aicher, Jacobs, and
  Clauset]{aicher2013adaptingARXIV}
Christopher Aicher, Abigail~Z. Jacobs, and Aaron Clauset.
\newblock Adapting the stochastic block model to edge-weighted networks.
\newblock \emph{ArXiv}, abs/1305.5782, 2013.

\bibitem[Aine et~al.(2017)Aine, Bockholt, Bustillo, Ca{\~n}ive, Caprihan,
  Gasparovic, Hanlon, Houck, Jung, and Lauriello]{aine2017multimodal}
CJ~Aine, Henry~Jeremy Bockholt, Juan~R Bustillo, Jos{\'e}~M Ca{\~n}ive, Arvind
  Caprihan, Charles Gasparovic, Faith~M. Hanlon, Jon~M. Houck, Rex~E. Jung, and
  John Lauriello.
\newblock Multimodal neuroimaging in schizophrenia: description and
  dissemination.
\newblock \emph{Neuroinformatics}, 15\penalty0 (4):\penalty0 343--364, 2017.
\newblock \doi{10.1007/s12021-017-9338-9}.
\newblock URL \url{https://www.ncbi.nlm.nih.gov/pmc/articles/PMC5671541/}.

\bibitem[Airoldi et~al.(2008)Airoldi, Blei, Fienberg, and
  Xing]{airoldi2008mixed}
Edoardo~M. Airoldi, David~M. Blei, Stephen~E. Fienberg, and Eric~P. Xing.
\newblock Mixed membership stochastic blockmodels.
\newblock \emph{Journal of Machine Learning Research}, 9\penalty0
  (65):\penalty0 1981--2014, 2008.
\newblock URL \url{http://jmlr.org/papers/v9/airoldi08a.html}.

\bibitem[Alexander-Bloch et~al.(2010)Alexander-Bloch, Gogtay, Meunier, Birn,
  Clasen, Lalonde, Lenroot, Giedd, and Bullmore]{alexanderbloch2010disrupted}
Aaron~F. Alexander-Bloch, Nitin Gogtay, David Meunier, Rasmus Birn, Liv Clasen,
  Francois Lalonde, Rhoshel Lenroot, Jay Giedd, and Edward~T. Bullmore.
\newblock Disrupted modularity and local connectivity of brain functional
  networks in childhood-onset schizophrenia.
\newblock \emph{Frontiers in Systems Neuroscience}, 4\penalty0 (147):\penalty0
  1--16, 2010.
\newblock ISSN 1662-5137.
\newblock \doi{10.3389/fnsys.2010.00147}.
\newblock URL \url{https://www.ncbi.nlm.nih.gov/pubmed/21031030}.

\bibitem[Athreya et~al.(2021)Athreya, Tang, Park, and
  Priebe]{athreya2021estimation}
Avanti Athreya, Minh Tang, Youngser Park, and Carey~E. Priebe.
\newblock On estimation and inference in latent structure random graphs.
\newblock \emph{Statistical Science}, 36\penalty0 (1):\penalty0 68--88, 2021.
\newblock \doi{10.1214/20-STS787}.
\newblock URL \url{https://doi.org/10.1214/20-STS787}.

\bibitem[Bazzi et~al.(2016)Bazzi, Porter, Williams, McDonald, Fenn, and
  Howison]{bazzi2016community}
Marya Bazzi, Mason~A. Porter, Stacy Williams, Mark McDonald, Daniel~J. Fenn,
  and Sam~D. Howison.
\newblock Community detection in temporal multilayer networks, with an
  application to correlation networks.
\newblock \emph{Multiscale Modeling \& Simulation}, 14\penalty0 (1):\penalty0
  1--41, 2016.
\newblock \doi{10.1137/15M1009615}.
\newblock URL \url{https://epubs.siam.org/doi/10.1137/15M1009615}.

\bibitem[Bhattacharyya and Chatterjee(2020)]{bhattacharyya2020generalARXIV}
Sharmodeep Bhattacharyya and Shirshendu Chatterjee.
\newblock General community detection with optimal recovery conditions for
  multi-relational sparse networks with dependent layers.
\newblock \emph{ArXiv}, abs/2004.03480, 2020.

\bibitem[Bickel and Chen(2009)]{bickel2009nonparametric}
Peter~J. Bickel and Aiyou Chen.
\newblock {A nonparametric view of network models and Newman--Girvan and other
  modularities}.
\newblock \emph{Proceedings of the National Academy of Sciences}, 106\penalty0
  (50):\penalty0 21068--21073, 2009.

\bibitem[Bickel et~al.(2013)Bickel, Choi, Chang, and
  Zhang]{bickel2013asymptotic}
Peter~J. Bickel, David Choi, Xiangyu Chang, and Hai Zhang.
\newblock {Asymptotic normality of maximum likelihood and its variational
  approximation for stochastic blockmodels}.
\newblock \emph{The Annals of Statistics}, 41\penalty0 (4):\penalty0
  1922--1943, 2013.
\newblock \doi{10.1214/13-AOS1124}.
\newblock URL \url{https://doi.org/10.1214/13-AOS1124}.

\bibitem[Blondel et~al.(2010)Blondel, Krings, and Thomas]{blondel2010regions}
Vincent Blondel, Gautier Krings, and Isabelle Thomas.
\newblock {Regions and borders of mobile telephony in Belgium and in the
  Brussels metropolitan zone}.
\newblock \emph{Brussels Studies}, 42:\penalty0 1--12, 10 2010.
\newblock \doi{10.4000/brussels.806}.
\newblock URL \url{https://journals.openedition.org/brussels/806}.

\bibitem[Blondel et~al.(2008)Blondel, Guillaume, Lambiotte, and
  Lefebvre]{blondel2008fast}
Vincent~D. Blondel, Jean-Loup Guillaume, Renaud Lambiotte, and Etienne
  Lefebvre.
\newblock Fast unfolding of communities in large networks.
\newblock \emph{Journal of Statistical Mechanics: Theory and Experiment},
  2008\penalty0 (10):\penalty0 10008, Oct 2008.
\newblock \doi{10.1088/1742-5468/2008/10/p10008}.
\newblock URL \url{https://doi.org/10.1088/1742-5468/2008/10/p10008}.

\bibitem[Bollob{\'{a}}s et~al.(2007)Bollob{\'{a}}s, Janson, and
  Riordan]{bollobas2007phase}
B{\'{e}}la Bollob{\'{a}}s, Svante Janson, and Oliver Riordan.
\newblock The phase transition in inhomogeneous random graphs.
\newblock \emph{Random Structures and Algorithms}, 31\penalty0 (1):\penalty0
  3--122, 2007.
\newblock \doi{10.1002/rsa.20168}.
\newblock URL \url{https://doi.org/10.1002%2Frsa.20168}.

\bibitem[Bordier et~al.(2017)Bordier, Nicolini, and Bifone]{bordier2017graph}
Cecile Bordier, Carlo Nicolini, and Angelo Bifone.
\newblock Graph analysis and modularity of brain functional connectivity
  networks: Searching for the optimal threshold.
\newblock \emph{Frontiers in Neuroscience}, 11\penalty0 (441):\penalty0 1--9,
  2017.
\newblock ISSN 1662-4548.
\newblock \doi{doi.org/10.3389/fnins.2017.00441}.
\newblock URL \url{https://www.ncbi.nlm.nih.gov/pubmed/28824364}.

\bibitem[Brandes et~al.(2008)Brandes, Delling, Gaertler, Gorke, Hoefer,
  Nikoloski, and Wagner]{brandes2008modularity}
Ulrik Brandes, Daniel Delling, Marco Gaertler, Robert Gorke, Martin Hoefer,
  Zoran Nikoloski, and Dorothea Wagner.
\newblock On modularity clustering.
\newblock \emph{IEEE Transactions on Knowledge and Data Engineering},
  20\penalty0 (2):\penalty0 172--188, 2008.
\newblock \doi{10.1109/TKDE.2007.190689}.
\newblock URL \url{https://ieeexplore.ieee.org/document/4358966}.

\bibitem[Bullmore and Sporns(2009)]{bullmore2009complex}
Ed~Bullmore and Olaf Sporns.
\newblock Complex brain networks: graph theoretical analysis of structural and
  functional systems.
\newblock \emph{Nature Reviews Neuroscience}, 10\penalty0 (3):\penalty0
  186--98, 2009.
\newblock \doi{10.1038/nrn2575}.
\newblock URL \url{https://www.ncbi.nlm.nih.gov/pubmed/19190637}.

\bibitem[Capriotti et~al.(2019)Capriotti, Ozturk, and
  Carter]{capriotti2019integrating}
Emidio Capriotti, Kivilcim Ozturk, and Hannah Carter.
\newblock Integrating molecular networks with genetic variant interpretation
  for precision medicine.
\newblock \emph{Wiley Interdisciplinary Reviews: Systems Biology and Medicine},
  11\penalty0 (3):\penalty0 e1443, 2019.
\newblock \doi{10.1002/wsbm.1443}.
\newblock URL \url{https://www.ncbi.nlm.nih.gov/pmc/articles/PMC6450710/}.

\bibitem[Chen et~al.(2021)Chen, Roch, Rohe, and Yu]{chen2021estimatingARXIV}
Fan Chen, Sebastien Roch, Karl Rohe, and Shuqi Yu.
\newblock Estimating graph dimension with cross-validated eigenvalues.
\newblock \emph{arXiv preprint arXiv:2108.03336}, 2021.

\bibitem[Chen and Lei(2018)]{chen2018network}
Kehui Chen and Jing Lei.
\newblock Network cross-validation for determining the number of communities in
  network data.
\newblock \emph{Journal of the American Statistical Association}, 113\penalty0
  (521):\penalty0 241--251, 2018.
\newblock \doi{10.1080/01621459.2016.1246365}.
\newblock URL
  \url{https://www.tandfonline.com/doi/full/10.1080/01621459.2016.1246365}.

\bibitem[Clauset et~al.(2004)Clauset, Newman, and Moore]{clauset2004finding}
Aaron Clauset, Mark E.~J. Newman, and Cristopher Moore.
\newblock Finding community structure in very large networks.
\newblock \emph{Physical Review E}, 70\penalty0 (6):\penalty0 066111, Dec 2004.
\newblock \doi{10.1103/PhysRevE.70.066111}.
\newblock URL \url{https://link.aps.org/doi/10.1103/PhysRevE.70.066111}.

\bibitem[Contreras et~al.(2019)Contreras, Avena-Koenigsberger, Risacher, West,
  Tallman, McDonald, Farlow, Apostolova, Goñi, Dzemidzic, Wu, Kessler, Jeub,
  Fortunato, Saykin, and Sporns]{contreras2019resting}
Joey~A. Contreras, Andrea Avena-Koenigsberger, Shannon~L. Risacher, John~D.
  West, Eileen Tallman, Brenna~C. McDonald, Martin~R. Farlow, Liana~G.
  Apostolova, Joaquín Goñi, Mario Dzemidzic, Yu-Chien Wu, Daniel Kessler,
  Lucas Jeub, Santo Fortunato, Andrew~J. Saykin, and Olaf Sporns.
\newblock Resting state network modularity along the prodromal late onset
  alzheimer's disease continuum.
\newblock \emph{NeuroImage: Clinical}, 22:\penalty0 101687, 2019.

\bibitem[De~Domenico(2017)]{dedominico2017multilayer}
Manlio De~Domenico.
\newblock Multilayer modeling and analysis of human brain networks.
\newblock \emph{Gigascience}, 6:\penalty0 1--8, 2017.
\newblock \doi{doi:10.1093/gigascience/gix004}.
\newblock URL \url{https://www.ncbi.nlm.nih.gov/pmc/articles/PMC5437946/}.

\bibitem[De~Domenico et~al.(2015{\natexlab{a}})De~Domenico, Lancichinetti,
  Arenas, and Rosvall]{dedomenico2015identifying}
Manlio De~Domenico, Andrea Lancichinetti, Alex Arenas, and Martin Rosvall.
\newblock Identifying modular flows on multilayer networks reveals highly
  overlapping organization in interconnected systems.
\newblock \emph{Physical Review X}, 5\penalty0 (1):\penalty0 011027, Mar
  2015{\natexlab{a}}.
\newblock \doi{10.1103/PhysRevX.5.011027}.
\newblock URL \url{https://link.aps.org/doi/10.1103/PhysRevX.5.011027}.

\bibitem[De~Domenico et~al.(2015{\natexlab{b}})De~Domenico, Nicosia, Arenas,
  and Latora]{dedominico2015structural}
Manlio De~Domenico, Vincenzo Nicosia, Alexandre Arenas, and Vito Latora.
\newblock Structural reducibility of multilayer networks.
\newblock \emph{Nature Communications}, 6\penalty0 (1):\penalty0 6864, Apr
  2015{\natexlab{b}}.
\newblock ISSN 2041-1723.
\newblock \doi{10.1038/ncomms7864}.
\newblock URL \url{https://doi.org/10.1038/ncomms7864}.

\bibitem[Dong et~al.(2012)Dong, Frossard, Vandergheynst, and
  Nefedov]{dong2012clustering}
Xiaowen Dong, Pascal Frossard, P.~Vandergheynst, and N.~Nefedov.
\newblock Clustering with multi-layer graphs: a spectral perspective.
\newblock \emph{IEEE Transactions on Signal Processing}, 60\penalty0
  (11):\penalty0 5820--5831, Nov 2012.
\newblock \doi{10.1109/tsp.2012.2212886}.
\newblock URL \url{http://dx.doi.org/10.1109/TSP.2012.2212886}.

\bibitem[Duch and Arenas(2005)]{duch2005community}
Jordi Duch and Alex Arenas.
\newblock Community detection in complex networks using extremal optimization.
\newblock \emph{Physical Review E}, 72:\penalty0 027104, Aug 2005.
\newblock \doi{10.1103/PhysRevE.72.027104}.
\newblock URL \url{https://link.aps.org/doi/10.1103/PhysRevE.72.027104}.

\bibitem[Erd\H{o}s and R\'{e}nyi(1959)]{erdos59random}
Paul Erd\H{o}s and Alfred R\'{e}nyi.
\newblock On random graphs i.
\newblock \emph{Publicationes Mathematicae Debrecen}, 6:\penalty0 290--297,
  1959.

\bibitem[Esfahlani et~al.(2021)Esfahlani, Jo, Puxeddu, Merritt, Tanner,
  Greenwell, Patel, Faskowitz, and Betzel]{esfahlani2021modularity}
Farnaz~Z. Esfahlani, Youngheun Jo, Maria~G. Puxeddu, Haily Merritt, Jacob~C.
  Tanner, Sarah Greenwell, Riya Patel, Joshua Faskowitz, and Richard~F. Betzel.
\newblock Modularity maximization as a flexible and generic framework for brain
  network exploratory analysis.
\newblock \emph{Neuroimage}, 244:\penalty0 118607, 2021.
\newblock ISSN 1095-9572.
\newblock \doi{10.1016/j.neuroimage.2021.118607}.
\newblock URL \url{https://www.ncbi.nlm.nih.gov/pubmed/34607022}.

\bibitem[Eulau and Siegel(1981)]{eulau1981social}
Heinz Eulau and Jonathan~W. Siegel.
\newblock Social network analysis and political behavior: A feasibility study.
\newblock \emph{The Western Political Quarterly}, 34\penalty0 (4):\penalty0
  499--509, 1981.
\newblock \doi{10.2307/447464}.
\newblock URL \url{http://www.jstor.org/stable/447464}.

\bibitem[Fan et~al.(2022)Fan, Fan, Han, and Lv]{fan2022asymptotic}
Jianqing Fan, Yingying Fan, Xiao Han, and Jinchi Lv.
\newblock Asymptotic theory of eigenvectors for random matrices with diverging
  spikes.
\newblock \emph{Journal of the American Statistical Association}, 117\penalty0
  (538):\penalty0 996--1009, 2022.

\bibitem[Fitzpatrick et~al.(2018)Fitzpatrick, Hobson, Mendelson, Rodríguez,
  Safran, Scordato, Servedio, Stern, Symes, and Kopp]{fitzpatrick2018theory}
Courtney~L. Fitzpatrick, Elizabeth~A. Hobson, Tamra~C. Mendelson, Rafael~L.
  Rodríguez, Rebecca~J. Safran, Elizabeth S.~C. Scordato, Maria~R. Servedio,
  Caitlin~A. Stern, Laurel~B. Symes, and Michael Kopp.
\newblock Theory meets empiry: A citation network analysis.
\newblock \emph{BioScience}, 68\penalty0 (10):\penalty0 805--812, 08 2018.
\newblock ISSN 0006-3568.
\newblock \doi{10.1093/biosci/biy083}.
\newblock URL \url{https://doi.org/10.1093/biosci/biy083}.

\bibitem[F{\"u}redi and Koml{\'o}s(1981)]{furedi1981eigenvalues}
Zolt{\'a}n F{\"u}redi and J{\'a}nos Koml{\'o}s.
\newblock The eigenvalues of random symmetric matrices.
\newblock \emph{Combinatorica}, 1:\penalty0 233--241, 1981.

\bibitem[Gilbert(1959)]{gilbert1959random}
Edgar~N. Gilbert.
\newblock {Random Graphs}.
\newblock \emph{The Annals of Mathematical Statistics}, 30\penalty0
  (4):\penalty0 1141 -- 1144, 1959.
\newblock \doi{10.1214/aoms/1177706098}.
\newblock URL \url{https://doi.org/10.1214/aoms/1177706098}.

\bibitem[Glasser et~al.(2016)Glasser, Coalson, Robinson, Hacker, Harwell,
  Yacoub, Ugurbil, Andersson, Beckmann, and Jenkinson]{glasser2016multi}
Matthew~F. Glasser, Timothy~S. Coalson, Emma~C. Robinson, Carl~D. Hacker, John
  Harwell, Essa Yacoub, Kamil Ugurbil, Jesper Andersson, Christian~F Beckmann,
  and Mark Jenkinson.
\newblock A multi-modal parcellation of human cerebral cortex.
\newblock \emph{Nature}, 536\penalty0 (7615):\penalty0 171--178, 2016.
\newblock URL \url{https://www.nature.com/articles/nature18933}.

\bibitem[Grayson and Fair(2017)]{grayson2017development}
David~S. Grayson and Damien~A. Fair.
\newblock Development of large-scale functional networks from birth to
  adulthood: a guide to the neuroimaging literature.
\newblock \emph{Neuroimage}, 160:\penalty0 15--31, 2017.
\newblock \doi{10.1016/j.neuroimage.2017.01.079}.
\newblock URL \url{https://www.ncbi.nlm.nih.gov/pmc/articles/PMC5538933/}.

\bibitem[Guime{r\`{a}} et~al.(2005)Guime{r\`{a}}, Mossa, Turtschi, and
  Amaral]{guimera2005worldwide}
R.~Guime{r\`{a}}, S.~Mossa, A.~Turtschi, and L.~A.~N. Amaral.
\newblock The worldwide air transportation network: anomalous centrality,
  community structure, and cities' global roles.
\newblock \emph{Proceedings of the National Academy of Sciences}, 102\penalty0
  (22):\penalty0 7794--7799, 2005.
\newblock \doi{10.1073/pnas.0407994102}.
\newblock URL \url{https://www.pnas.org/doi/abs/10.1073/pnas.0407994102}.

\bibitem[Han et~al.(2023)Han, Yang, and Fan]{han2023universal}
Xiao Han, Qing Yang, and Yingying Fan.
\newblock {Universal rank inference via residual subsampling with application
  to large networks}.
\newblock \emph{The Annals of Statistics}, 51\penalty0 (3):\penalty0
  1109--1133, 2023.
\newblock \doi{10.1214/23-AOS2282}.
\newblock URL \url{https://doi.org/10.1214/23-AOS2282}.

\bibitem[Holland et~al.(1983)Holland, Laskey, and
  Leinhardt]{holland1983stochastic}
Paul~W. Holland, Kathryn~Blackmond Laskey, and Samuel Leinhardt.
\newblock Stochastic blockmodels: first steps.
\newblock \emph{Social Networks}, 5\penalty0 (2):\penalty0 109--137, 1983.
\newblock ISSN 0378-8733.
\newblock \doi{https://doi.org/10.1016/0378-8733(83)90021-7}.
\newblock URL
  \url{https://www.Elsevier.com/science/article/pii/0378873383900217}.

\bibitem[Hubert and Arabie(1985)]{hubert1985comparing}
Lawrence Hubert and Phipps Arabie.
\newblock Comparing partitions.
\newblock \emph{Journal of Classification}, 2\penalty0 (1):\penalty0 193--218,
  12 1985.
\newblock \doi{10.1007/BF01908075}.
\newblock URL \url{https://doi.org/10.1007/BF01908075}.

\bibitem[Hwang et~al.(2023)Hwang, Xu, Chatterjee, and
  Bhattacharyya]{hwang2023estimation}
Neil Hwang, Jiarui Xu, Shirshendu Chatterjee, and Sharmodeep Bhattacharyya.
\newblock On the estimation of the number of communities for sparse networks.
\newblock \emph{Journal of the American Statistical Association}, pages 1--16,
  2023.
\newblock \doi{10.1080/01621459.2023.2223793}.
\newblock URL \url{https://doi.org/10.1080/01621459.2023.2223793}.

\bibitem[Jin et~al.(2023)Jin, Ke, Luo, and Wang]{jin2023optimal}
Jiashun Jin, Zheng~Tracy Ke, Shengming Luo, and Minzhe Wang.
\newblock Optimal estimation of the number of network communities.
\newblock \emph{Journal of the American Statistical Association}, 118\penalty0
  (543):\penalty0 2101--2116, 2023.
\newblock \doi{10.1080/01621459.2022.2035736}.
\newblock URL \url{https://doi.org/10.1080/01621459.2022.2035736}.

\bibitem[Karrer and Newman(2011)]{karrer2011stochastic}
Brian Karrer and Mark E.~J. Newman.
\newblock Stochastic blockmodels and community structure in networks.
\newblock \emph{Physical Review E}, 83\penalty0 (1):\penalty0 016107, Jan 2011.
\newblock \doi{10.1103/PhysRevE.83.016107}.
\newblock URL \url{https://link.aps.org/doi/10.1103/PhysRevE.83.016107}.

\bibitem[Klimm et~al.(2022)Klimm, Jones, and Schaub]{klimm2021modularity}
Florian Klimm, Nick~S. Jones, and Michael~T. Schaub.
\newblock Modularity maximization for graphons.
\newblock \emph{SIAM Journal on Applied Mathematics}, 82\penalty0 (6):\penalty0
  1930--1952, 2022.
\newblock \doi{10.1137/22M1492003}.
\newblock URL \url{https://doi.org/10.1137/22M1492003}.

\bibitem[Koo et~al.(2023)Koo, Tang, and Trosset]{koo2023popularity}
John Koo, Minh Tang, and Michael~W. Trosset.
\newblock Popularity adjusted block models are generalized random dot product
  graphs.
\newblock \emph{Journal of Computational and Graphical Statistics}, 32\penalty0
  (1):\penalty0 131--144, 2023.
\newblock \doi{10.1080/10618600.2022.2081576}.
\newblock URL \url{https://doi.org/10.1080/10618600.2022.2081576}.

\bibitem[Lee and Wilkinson(2019)]{lee2019review}
Clement Lee and Darren~J. Wilkinson.
\newblock A review of stochastic block models and extensions for graph
  clustering.
\newblock \emph{Applied Network Science}, 4\penalty0 (1):\penalty0 1--50, Dec
  2019.
\newblock \doi{10.1007/s41109-019-0232-2}.
\newblock URL
  \url{https://appliednetsci.springeropen.com/articles/10.1007/s41109-019-0232-2}.

\bibitem[Lei et~al.(2022)Lei, Qin, Pinaya, Young, Van~Amelsvoort, Marcelis,
  Donohoe, Mothersill, Corvin, Vieira, Lui, Scarpazza, Arango, Bullmore, Gong,
  McGuire, and Mechelli]{lei2022graph}
Du~Lei, Kun Qin, Walter H.~L. Pinaya, Jonathan Young, Therese Van~Amelsvoort,
  Machteld Marcelis, Gary Donohoe, David~O. Mothersill, Aiden Corvin, Sandra
  Vieira, Su~Lui, Cristina Scarpazza, Celso Arango, Ed~Bullmore, Qiyong Gong,
  Philip McGuire, and Andrea Mechelli.
\newblock Graph convolutional networks reveal network-level functional
  dysconnectivity in schizophrenia.
\newblock \emph{Schizophrenia Bulletin}, 48\penalty0 (4):\penalty0 881--892, 05
  2022.
\newblock \doi{10.1093/schbul/sbac047}.
\newblock URL \url{https://doi.org/10.1093/schbul/sbac047}.

\bibitem[Lei(2016)]{lei2016goodnessoffit}
Jing Lei.
\newblock {A goodness-of-fit test for stochastic block models}.
\newblock \emph{The Annals of Statistics}, 44\penalty0 (1):\penalty0 401--424,
  2016.
\newblock \doi{10.1214/15-AOS1370}.
\newblock URL \url{https://doi.org/10.1214/15-AOS1370}.

\bibitem[Lei et~al.(2019)Lei, Chen, and Lynch]{lei2019consistent}
Jing Lei, Kehui Chen, and Brian Lynch.
\newblock Consistent community detection in multi-layer network data.
\newblock \emph{Biometrika}, 107\penalty0 (1):\penalty0 61--73, 2019.
\newblock \doi{10.1093/biomet/asz068}.
\newblock URL \url{https://doi.org/10.1093/biomet/asz068}.

\bibitem[Leung et~al.(2014)Leung, Tanbeer, and Cameron]{leung2014interactive}
Carson Kai-Sang Leung, Syed~K. Tanbeer, and Juan~J. Cameron.
\newblock Interactive discovery of influential friends from social networks.
\newblock \emph{Social Network Analysis and Mining}, 4\penalty0 (1):\penalty0
  1--13, 02 2014.
\newblock ISSN 1869-5469.
\newblock \doi{10.1007/s13278-014-0154-z}.
\newblock URL
  \url{https://link.springer.com/article/10.1007/s13278-014-0154-z}.

\bibitem[Li et~al.(2020)Li, Levina, and Zhu]{tianxi2020network}
Tianxi Li, Elizaveta Levina, and Ji~Zhu.
\newblock Network cross-validation by edge sampling.
\newblock \emph{Biometrika}, 107\penalty0 (2):\penalty0 257--276, 04 2020.
\newblock ISSN 0006-3444.
\newblock \doi{10.1093/biomet/asaa006}.
\newblock URL \url{https://doi.org/10.1093/biomet/asaa006}.

\bibitem[Li and Qi(2020)]{li2020asymptotic}
Yang Li and Yongcheng Qi.
\newblock Asymptotic distribution of modularity in networks.
\newblock \emph{Metrika}, 83\penalty0 (4):\penalty0 467--484, May 2020.
\newblock ISSN 0026-1335.
\newblock \doi{10.1007/s00184-019-00740-7}.
\newblock URL
  \url{https://link.springer.com/content/pdf/10.1007/s00184-019-00740-7.pdf}.

\bibitem[Lyzinski et~al.(2014)Lyzinski, Sussman, Tang, Athreya, and
  Priebe]{lyzinski2014perfect}
Vince Lyzinski, Daniel~L. Sussman, Minh Tang, Avanti Athreya, and Carey~E.
  Priebe.
\newblock Perfect clustering for stochastic blockmodel graphs via adjacency
  spectral embedding.
\newblock \emph{Electronic Journal of Statistics}, 8\penalty0 (2):\penalty0
  2905--2922, 2014.
\newblock \doi{10.1214/14-EJS978}.
\newblock URL \url{https://doi.org/10.1214/14-EJS978}.

\bibitem[Ma and Barnett(2020)]{ma2020theasymptotic}
Rong Ma and Ian Barnett.
\newblock The asymptotic distribution of modularity in weighted signed
  networks.
\newblock \emph{Biometrika}, 108\penalty0 (1):\penalty0 1--16, Jul 2020.
\newblock \doi{10.1093/biomet/asaa059}.
\newblock URL \url{https://www.ncbi.nlm.nih.gov/pmc/articles/PMC8300091/}.

\bibitem[Magnus and Neudecker(1980)]{magnus1980elimination}
Jan~R. Magnus and Heinz Neudecker.
\newblock The elimination matrix: Some lemmas and applications.
\newblock \emph{SIAM Journal on Algebraic Discrete Methods}, 1\penalty0
  (4):\penalty0 422--449, 1980.
\newblock \doi{10.1137/0601049}.
\newblock URL \url{https://doi.org/10.1137/0601049}.

\bibitem[Mucha et~al.(2010)Mucha, Richardson, Macon, Porter, and
  Onnela]{mucha2010community}
Peter~J. Mucha, Thomas Richardson, Kevin Macon, Mason~A. Porter, and
  Jukka-Pekka Onnela.
\newblock Community structure in time-dependent, multiscale, and multiplex
  networks.
\newblock \emph{Science}, 328\penalty0 (5980):\penalty0 876--878, May 2010.
\newblock ISSN 1095-9203.
\newblock \doi{10.1126/science.1184819}.
\newblock URL \url{http://dx.doi.org/10.1126/science.1184819}.

\bibitem[Newman(2004)]{newman2004fast}
Mark E.~J. Newman.
\newblock Fast algorithm for detecting community structure in networks.
\newblock \emph{Physical Review E}, 69:\penalty0 066133, Jun 2004.
\newblock \doi{10.1103/PhysRevE.69.066133}.
\newblock URL \url{https://link.aps.org/doi/10.1103/PhysRevE.69.066133}.

\bibitem[Newman(2006{\natexlab{a}})]{newman2006finding}
Mark E.~J. Newman.
\newblock Finding community structure in networks using the eigenvectors of
  matrices.
\newblock \emph{Physical Review E}, 74\penalty0 (3):\penalty0 036104, Sep
  2006{\natexlab{a}}.
\newblock \doi{10.1103/PhysRevE.74.036104}.
\newblock URL \url{https://link.aps.org/doi/10.1103/PhysRevE.74.036104}.

\bibitem[Newman(2006{\natexlab{b}})]{newman2006modularity}
Mark E.~J. Newman.
\newblock Modularity and community structure in networks.
\newblock \emph{Proceedings of the National Academy of Sciences}, 103\penalty0
  (23):\penalty0 8577--8582, 2006{\natexlab{b}}.
\newblock \doi{10.1073/pnas.0601602103}.
\newblock URL \url{https://www.pnas.org/content/103/23/8577.full.pdf}.

\bibitem[Newman(2018)]{newman2018networks}
Mark E.~J. Newman.
\newblock \emph{Networks}.
\newblock Oxford University Press, 2 edition, 2018.
\newblock \doi{10.1093/oso/9780198805090.001.0001}.

\bibitem[Newman and Girvan(2004)]{newman2004finding}
Mark E.~J. Newman and Michelle Girvan.
\newblock Finding and evaluating community structure in networks.
\newblock \emph{Physical Review E}, 69\penalty0 (2):\penalty0 026113, Feb 2004.
\newblock \doi{10.1103/PhysRevE.69.026113}.
\newblock URL \url{https://link.aps.org/doi/10.1103/PhysRevE.69.026113}.

\bibitem[Porter et~al.(2005)Porter, Mucha, Newman, and
  Warmbrand]{porter2005network}
Mason~A. Porter, Peter~J. Mucha, Mark E.~J. Newman, and Casey~M. Warmbrand.
\newblock {A network analysis of committees in the U.S. House of
  Representatives}.
\newblock \emph{Proceedings of the National Academy of Sciences}, 102\penalty0
  (20):\penalty0 7057--7062, 2005.
\newblock \doi{10.1073/pnas.0500191102}.
\newblock URL \url{https://www.pnas.org/doi/abs/10.1073/pnas.0500191102}.

\bibitem[Power et~al.(2011)Power, Cohen, Nelson, Wig, Barnes, Church, Vogel,
  Laumann, Miezin, Schlaggar, and Petersen]{power2011functional}
Jonathan~D. Power, Alexander~L. Cohen, Steven~M. Nelson, Gagan~S. Wig,
  Kelly~Anne Barnes, Jessica~A. Church, Alecia~C. Vogel, Timothy~O. Laumann,
  Fran~M. Miezin, Bradley~L. Schlaggar, and Steven~E. Petersen.
\newblock Functional network organization of the human brain.
\newblock \emph{Neuron}, 72\penalty0 (4):\penalty0 665--678, 2011.
\newblock \doi{10.1016/j.neuron.2011.09.006}.
\newblock URL \url{https://www.ncbi.nlm.nih.gov/pmc/articles/PMC3222858/}.

\bibitem[Priebe et~al.(2019)Priebe, Park, Vogelstein, Conroy, Lyzinski, Tang,
  Athreya, Cape, and Bridgeford]{priebe2019twotruths}
Carey~E. Priebe, Youngser Park, Joshua~T. Vogelstein, John~M. Conroy, Vince
  Lyzinski, Minh Tang, Avanti Athreya, Joshua Cape, and Eric Bridgeford.
\newblock On a two-truths phenomenon in spectral graph clustering.
\newblock \emph{Proceedings of the National Academy of Sciences}, 116\penalty0
  (13):\penalty0 5995--6000, 2019.
\newblock ISSN 0027-8424.
\newblock \doi{10.1073/pnas.1814462116}.
\newblock URL \url{https://www.pnas.org/content/116/13/5995}.

\bibitem[Reichardt and Bornholdt(2006)]{reichardt2006statistical}
J\"org Reichardt and Stefan Bornholdt.
\newblock Statistical mechanics of community detection.
\newblock \emph{Physical Review E}, 74:\penalty0 016110, Jul 2006.
\newblock \doi{10.1103/PhysRevE.74.016110}.
\newblock URL \url{https://link.aps.org/doi/10.1103/PhysRevE.74.016110}.

\bibitem[Reli{\'{o}}n et~al.(2019)Reli{\'{o}}n, Kessler, Levina, and
  Taylor]{arroyo2019network}
Jes{\'{u}}s D.~Arroyo Reli{\'{o}}n, Daniel Kessler, Elizaveta Levina, and
  Stephan~F. Taylor.
\newblock Network classification with applications to brain connectomics.
\newblock \emph{The Annals of Applied Statistics}, 13\penalty0 (3):\penalty0
  1648--1677, Sep 2019.
\newblock \doi{10.1214/19-aoas1252}.
\newblock URL \url{https://doi.org/10.1214%2F19-aoas1252}.

\bibitem[Rubin-Delanchy et~al.(2022)Rubin-Delanchy, Cape, Tang, and
  Priebe]{rubindelanchy2022astatistical}
Patrick Rubin-Delanchy, Joshua Cape, Minh Tang, and Carey~E. Priebe.
\newblock {A statistical interpretation of spectral embedding: the generalised
  random dot product graph}.
\newblock \emph{Journal of the Royal Statistical Society Series B}, 84\penalty0
  (4):\penalty0 1446--1473, September 2022.
\newblock \doi{10.1111/rssb.12509}.
\newblock URL \url{https://academic.oup.com/jrsssb/article/84/4/1446/7073272}.

\bibitem[Rubinov and Sporns(2010)]{rubinov2010complex}
Mikail Rubinov and Olaf Sporns.
\newblock Complex network measures of brain connectivity: uses and
  interpretations.
\newblock \emph{Neuroimage}, 52\penalty0 (3):\penalty0 1059--1069, 2010.
\newblock \doi{10.1016/j.neuroimage.2009.10.003}.
\newblock URL
  \url{https://www.Elsevier.com/science/article/pii/S105381190901074X?via%3Dihub}.

\bibitem[Sengupta and Chen(2018)]{sengupta2018blockmodel}
Srijan Sengupta and Yuguo Chen.
\newblock {A block model for node popularity in networks with community
  structure}.
\newblock \emph{Journal of the Royal Statistical Society Series B}, 80\penalty0
  (2):\penalty0 365--386, March 2018.
\newblock \doi{10.1111/rssb.12245}.
\newblock URL
  \url{https://ideas.repec.org/a/bla/jorssb/v80y2018i2p365-386.html}.

\bibitem[Sporns and Betzel(2016)]{sporns2016modular}
Olaf Sporns and Richard~F. Betzel.
\newblock Modular brain networks.
\newblock \emph{Annual Review of Psychology}, 67:\penalty0 1--30, 2016.
\newblock ISSN 1545-2085.
\newblock \doi{10.1146/annurev-psych-122414-033634}.
\newblock URL \url{https://www.ncbi.nlm.nih.gov/pubmed/26393868}.

\bibitem[Sweet(2015)]{sweet2015incorporating}
Tracy~M. Sweet.
\newblock Incorporating covariates into stochastic blockmodels.
\newblock \emph{Journal of Educational and Behavioral Statistics}, 40\penalty0
  (6):\penalty0 635--664, 2015.
\newblock \doi{10.3102/1076998615606110}.
\newblock URL \url{https://journals.sagepub.com/doi/10.3102/1076998615606110}.

\bibitem[Tan et~al.(2019)Tan, Huang, Zhang, and Li]{tan2019network}
Aidi Tan, Huiya Huang, Peng Zhang, and Shao Li.
\newblock Network-based cancer precision medicine: a new emerging paradigm.
\newblock \emph{Cancer Letters}, 458:\penalty0 39--45, 2019.
\newblock \doi{10.1016/j.canlet.2019.05.015}.
\newblock URL
  \url{https://www.Elsevier.com/science/article/pii/S0304383519303106?via%3Dihub}.

\bibitem[Tang et~al.(2022)Tang, Cape, and Priebe]{tang2022asymptotically}
Minh Tang, Joshua Cape, and Carey~E. Priebe.
\newblock {Asymptotically efficient estimators for stochastic blockmodels: the
  naive MLE, the rank-constrained MLE, and the spectral estimator}.
\newblock \emph{Bernoulli}, 28\penalty0 (2):\penalty0 1049--1073, 2022.
\newblock \doi{10.3150/21-BEJ1376}.
\newblock URL \url{https://doi.org/10.3150/21-BEJ1376}.

\bibitem[Traag et~al.(2019)Traag, Waltman, and van Eck]{traag2019fromlouvain}
Vincent~A. Traag, Ludo Waltman, and Nees~Jan van Eck.
\newblock {From Louvain to Leiden: guaranteeing well-connected communities}.
\newblock \emph{Scientific Reports}, 9\penalty0 (1):\penalty0 5233, Mar 2019.
\newblock \doi{10.1038/s41598-019-41695-z}.
\newblock URL \url{https://www.nature.com/articles/s41598-019-41695-z}.

\bibitem[van~den Heuvel and Fornito(2014)]{van2014brain}
Martijn~P. van~den Heuvel and Alex Fornito.
\newblock Brain networks in schizophrenia.
\newblock \emph{Neuropsychology Review}, 24:\penalty0 32--48, 2014.

\bibitem[von Luxburg(2007)]{luxburg2007tutorial}
Ulrike von Luxburg.
\newblock A tutorial on spectral clustering.
\newblock \emph{Statistics and Computing}, 17\penalty0 (4):\penalty0 395--416,
  Dec 2007.
\newblock \doi{10.1007/s11222-007-9033-z}.
\newblock URL \url{https://doi.org/10.1007/s11222-007-9033-z}.

\bibitem[Xie(2022)]{xie2022entrywise}
Fangzheng Xie.
\newblock Entrywise limit theorems of eigenvectors for signal-plus-noise matrix
  models with weak signals.
\newblock \emph{Bernoulli}, 30\penalty0 (1):\penalty0 388--418, 2022.
\newblock \doi{10.3150/23-BEJ1602}.
\newblock URL \url{https://doi.org/10.3150/23-BEJ1602}.

\bibitem[Xie and Xu(2023)]{xie2023efficient}
Fangzheng Xie and Yanxun Xu.
\newblock Efficient estimation for random dot product graphs via a one-step
  procedure.
\newblock \emph{Journal of the American Statistical Association}, 118\penalty0
  (541):\penalty0 651--664, 2023.
\newblock \doi{10.1080/01621459.2021.1948419}.
\newblock URL \url{https://doi.org/10.1080/01621459.2021.1948419}.

\bibitem[Zachary(1977)]{zachary1977aninformation}
Wayne~W. Zachary.
\newblock An information flow model for conflict and fission in small groups.
\newblock \emph{Journal of Anthropological Research}, 33\penalty0 (4):\penalty0
  452--473, 1977.
\newblock ISSN 00917710.
\newblock \doi{10.1086/jar.33.4.3629752}.
\newblock URL \url{http://www.jstor.org/stable/3629752}.

\bibitem[Zhang et~al.(2022)Zhang, Guo, and Chang]{zhang2022randomized}
Hai Zhang, Xiao Guo, and Xiangyu Chang.
\newblock Randomized spectral clustering in large-scale stochastic block
  models.
\newblock \emph{Journal of Computational and Graphical Statistics}, 31\penalty0
  (3):\penalty0 887--906, 2022.
\newblock \doi{10.1080/10618600.2022.2034636}.
\newblock URL
  \url{https://www.tandfonline.com/doi/abs/10.1080/10618600.2022.2034636}.

\bibitem[Zhang and Chen(2017)]{zhang2017hypothesis}
Jingfei Zhang and Yuguo Chen.
\newblock A hypothesis testing framework for modularity based network community
  detection.
\newblock \emph{Statistica Sinica}, 27\penalty0 (1):\penalty0 437--456, 2017.
\newblock ISSN 1545-2085.
\newblock \doi{10.5705/ss.202015.0040}.
\newblock URL \url{http://www.jstor.org/stable/44114379}.

\bibitem[Zhang et~al.(2021)Zhang, Jiang, Qiao, and Liu]{zhang2021modularity}
Yangyang Zhang, Xiao Jiang, Lishan Qiao, and Mingxia Liu.
\newblock Modularity-guided functional brain network analysis for early-stage
  dementia identification.
\newblock \emph{Frontiers in Neuroscience}, 15:\penalty0 720909, 2021.

\bibitem[Zhang et~al.(2019)Zhang, Chen, Sampson, Hwang, and
  Luna]{zhang2019node}
Yun Zhang, Kehui Chen, Allan Sampson, Kai Hwang, and Beatriz Luna.
\newblock Node features adjusted stochastic block model.
\newblock \emph{Journal of Computational and Graphical Statistics}, 28\penalty0
  (2):\penalty0 362--373, 2019.
\newblock \doi{10.1080/10618600.2018.1530117}.
\newblock URL
  \url{https://www.tandfonline.com/doi/full/10.1080/10618600.2018.1530117}.

\bibitem[Zhu and Ghodsi(2006)]{zhu2006automatic}
Mu~Zhu and Ali Ghodsi.
\newblock Automatic dimensionality selection from the scree plot via the use of
  profile likelihood.
\newblock \emph{Computational Statistics \& Data Analysis}, 51\penalty0
  (2):\penalty0 918--930, 2006.
\newblock ISSN 0167-9473.
\newblock \doi{https://doi.org/10.1016/j.csda.2005.09.010}.
\newblock URL
  \url{https://www.Elsevier.com/science/article/pii/S0167947305002343}.

\end{thebibliography}

\clearpage
\appendix
\setcounter{page}{1}
\section{Appendix to ``On inference for modularity statistics in structured networks'' by Anirban Mitra, Konasale Prasad, and Joshua Cape}
\label{sec:suppl}
This document contains proofs of the stated theorems, additional simulation examples, and further discussion pertaining to the real data analysis.

\subsection{Proofs of asymptotic normality}
\label{sec:suppl:proofs}

\begin{lemma}[\cite{bickel2013asymptotic}]\label{lemma:BL_normality}
	For $n \ge 1$, suppose $\mathbf{A}^{(n)} \sim \operatorname{SBM}(\mathbf{B}, \boldsymbol{\pi})$ is a sequence of stochastic blockmodel graphs with sparsity factor $\rho_{n}$ satisfying $n\rho_{n} = \omega(\log n)$. Let $\mathbf{B}^{(\Like)} = (B_{kl}^{(\Like)})$ be as in \cref{def_block_estimate}. Then, as $n \rightarrow \infty$,
	\begin{equation}
		n\rho_{n}^{1/2}\operatorname{vech}(\widehat{\mathbf{B}}^{(\Like)} - \mathbf{B}) 
		\xrightarrow{\operatorname{d}}
		\mathscr{N}(\mathbf{0}, \mathbf{D}^{-1}). \label{dist1}
	\end{equation}
	The diagonal matrix $\mathbf{D} = (\mathbf{D}^{-1})^{-1}$ is defined in \cref{D-form} of the main text.
\end{lemma}

\begin{lemma}[\cite{tang2022asymptotically}]\label{lemma:BS_normality}
	For $n \ge 1$, suppose $\mathbf{A}^{(n)} \sim \operatorname{SBM}(\mathbf{B}, \boldsymbol{\pi})$ is a sequence of stochastic blockmodel graphs with sparsity factor $\rho_{n}$ satisfying $n\rho_{n} = \omega(\sqrt{n})$. Let $\widehat{\mathbf{B}}^{(\Spec)} = (\widehat{B}_{kl}^{(\Spec)})$ be as in \cref{def_block_estimate}. Then, as $n\rightarrow\infty$,
\begin{equation}
    n\rho_{n}^{1/2}\operatorname{vech}(\widehat{\mathbf{B}}^{(\Spec)} - \mathbf{B}) - \rho_{n}^{-1/2}\operatorname{vech}(\mathbf{\Theta}) 
    \xrightarrow{\operatorname{d}} 
    \mathscr{N}(\mathbf{0}, \widetilde{\mathbf{\Gamma}}). \label{dist2}
\end{equation}
Here, $\boldsymbol{\Theta}$ per \cref{theta_eqs} is the asymptotic bias and $\widetilde{\mathbf{\Gamma}}$ per \cref{gamma_tilde} is the asymptotic covariance matrix.
\end{lemma}

Importantly, \cite{tang2022asymptotically} establishes the decomposition
\begin{align}\label{eq:B_diff_normality}             
    & n\rho_{n}^{1/2}\operatorname{vech}\left(\widehat{\mathbf{B}}^{(\Spec)} - \mathbf{B} - (n\rho_{n})^{-1}\mathbf{\Theta}\right) \nonumber\\
    &\hspace{1em}=
    n\rho_{n}^{1/2} \mathcal{L}_{K}(\mathbf{I} - \Breve{\mathbf{\Pi}}_{\mathbf{V}}^{\perp} \otimes \Breve{\mathbf{\Pi}}_{\mathbf{V}}^{\perp})\mathcal{D}_{K} \operatorname{vech}(\widehat{\mathbf{B}}^{(\Like)} - \mathbf{B}) + O_{\mathbb{P}}\left(n^{-1/2}\rho_{n}^{-1}\right).
\end{align}
In \cref{eq:B_diff_normality}, $\Breve{\mathbf{\Pi}}_{\mathbf{V}}^{\perp} = \mathbf{I} - \mathbf{\mathbf{V}} (\mathbf{V}^{\top} \mathbf{Z}^{\top}\mathbf{Z} \mathbf{V})^{-1}\mathbf{V}^{\top} \mathbf{Z}^{\top}\mathbf{Z}$ and therefore by the law of large numbers $\Breve{\mathbf{\Pi}}_{\mathbf{V}}^{\perp} \xrightarrow{\operatorname{a.s.}} \widetilde{\mathbf{\Pi}}_{\mathbf{V}}^{\perp}$ in the large-network limit. When $\mathbf{B}$ is full rank, then necessarily $\mathbf{\Theta} = \mathbf{0}$ and the limiting covariance matrix for \cref{dist2} agrees with $\mathbf{D}^{-1}$. The above decomposition yields the following corollary in the form of a lemma.

\begin{lemma}[\cite{tang2022asymptotically}]\label{lemma:B_diff_normality}
	For $n \ge 1$, suppose $\mathbf{A}^{(n)} \sim \operatorname{SBM}(\mathbf{B}, \boldsymbol{\pi})$ is a sequence of stochastic blockmodel graphs with sparsity factor $\rho_{n}$ satisfying $n\rho_{n} = \omega(\sqrt{n})$. Provided $\mathbf{B}$ is strictly rank deficient, then as $n \rightarrow \infty$,
\begin{align}
    & n\rho_{n}^{1/2}\operatorname{vech}(\widehat{\mathbf{B}}^{(\Spec)} - \widehat{\mathbf{B}}^{(\Like)})
    -
    \rho_{n}^{-1/2}\operatorname{vech}(\mathbf{\Theta}) 
    \xrightarrow{\operatorname{d}}
    \mathscr{N}(\mathbf{0}, \mathbf{\Gamma}). \label{dist3}
\end{align}
In \cref{dist3}, $\boldsymbol{\Gamma} = \mathcal{L}_{K}(\widetilde{\mathbf{\Pi}}_{\mathbf{V}}^{\perp} \otimes \widetilde{\mathbf{\Pi}}_{\mathbf{V}}^{\perp})\mathcal{D}_{K} \mathbf{D}^{-1}\mathcal{D}_{K}^{\top} (\widetilde{\mathbf{\Pi}}_{\mathbf{V}}^{\perp} \otimes \widetilde{\mathbf{\Pi}}_{\mathbf{V}}^{\perp})^{\top}\mathcal{L}_{K}^{\top}$.
\end{lemma}

The aforementioned lemmas facilitate proving the main theorems.

\begin{proof}[Proofs of \cref{thrm:mod_like,thrm:mod_spec,thrm:mod_res}]
	Recall the proposed modularity functions in \cref{def_mods3}. Due to the associated block structure and use of $\boldsymbol{\tau}$, it holds that
\begin{equation*}
	\sum_{i,j = 1}^{n}\left(A_{ij} - P_{ij}\right)\mathbb{I}_{\{\tau_{i}=\tau_{j}\}}
    =
    \sum_{k = 1}^{K}n_{k}^{2}\rho_{n}\left(\widehat{{B}}_{kk}^{(\Like)} - {B}_{kk}\right),
\end{equation*}
\begin{equation*}
	\sum_{i,j = 1}^{n}\left(\widehat{{A}}_{ij} - P_{ij}\right)\mathbb{I}_{\{\tau_{i}=\tau_{j}\}}
    =
    \sum_{k = 1}^{K}n_{k}^{2}\rho_{n}\left(\widehat{{B}}_{kk}^{(\Spec)} - {B}_{kk}\right),
\end{equation*}
and
\begin{equation*}
	\sum_{i,j = 1}^{n}\left({A}_{ij} - \widehat{A}_{ij}\right)\mathbb{I}_{\{\tau_{i}=\tau_{j}\}}
    =
    \sum_{i,j = 1}^{n}\left(A_{ij} - P_{ij} -  \widehat{A}_{ij} + P_{ij}\right)\mathbb{I}_{\{\tau_{i}=\tau_{j}\}}
     =
     \sum_{k = 1}^{K}n_{k}^{2}\rho_{n}\left(\widehat{{B}}_{kk}^{(\Like)} - \widehat{{B}}_{kk}^{(\Spec)}\right).
\end{equation*}
Consequently, the stated asymptotic normality of $Q_{\Like}$, $Q_{\Spec}$, and $Q_{\Res}$ each hold by a direct application of \cref{lemma:BL_normality,lemma:BS_normality,lemma:B_diff_normality}. 
\end{proof}

\subsubsection{Comment on the influence of matrix rank}
\label{sec:suppl:rank_discussion}

Here, we address the question raised by a referee ``How robust is the asymptotic normality result for residual-based modularity if $\mathbf{B}$ is only approximately low-rank?''

In fact, there are some technical subtleties when dealing with $Q_{\Res}$ for full-rank connectivity matrices $\mathbf{B}$. Below, we highlight why it is anticipated that the scaling of $Q_{\Res}$ in the full-rank setting will differ from the common scaling for $Q_{\Like}$ and $Q_{\Spec}$ in our main theorems.

For simplicity, fix the sparsity factor $\rho_{n} \equiv 1$. Write $\mathbf{A} \sim \operatorname{ER}(p)$ to indicate the adjacency matrix for a loopy undirected Erd\H{o}s--R\'{e}nyi random graph on $n$ nodes with edge probability $p$. Note that ER graphs represent $K=1$ stochastic blockmodel graphs that are necessarily full rank with $B_{11} = p$ and $d=1$. Let $\widehat{\mathbf{A}}$ denote the rank-one spectral truncation of $\mathbf{A}$. Let $\mathbf{1}$ denote the vector of all ones, which corresponds to the ER ground-truth partition vector. For this setting, the residual variant of modularity is given by
\begin{align*}
	Q_{\Res}
	&=
	\sum_{i,j = 1}^{n}\left({A}_{ij} - \widehat{{A}}_{ij}\right)\mathbb{I}_{\{\tau_{i}=\tau_{j}\}}
	\qquad\qquad \text{(general definition)}\\
	&=
	\boldsymbol{1}^{\top}(\mathbf{A} - \widehat{\mathbf{A}})\mathbf{1}.
	\qquad\qquad \text{(special form in rank one setting)}
\end{align*}
Write the rank-one spectral approximation of $\mathbf{A}$ as $\widehat{\mathbf{A}} = \widehat{\lambda}\widehat{\mathbf{u}}\widehat{\mathbf{u}}^{\top}$ and similarly write $\mathbb{E}[\mathbf{A}] = p \boldsymbol{1} \boldsymbol{1}^{\top} = \lambda \mathbf{u} \mathbf{u}^{\top}$, where $\lambda = np$ and $\mathbf{u}$ is the constant vector with entries $n^{-1/2}$. Let $\langle \cdot, \cdot \rangle$ denote shorthand for the Euclidean dot product. Consider the expansion of $Q_{\Res}$ given by
\begin{align*}
	\boldsymbol{1}^{\top}(\mathbf{A} - \widehat{\mathbf{A}})\boldsymbol{1}
	&=
	\langle \mathbf{A} \boldsymbol{1}, \boldsymbol{1} \rangle - \widehat{\lambda} \times |\langle \widehat{\mathbf{u}}, \mathbf{u}\rangle|^{2} \times n.
\end{align*}
\cite{furedi1981eigenvalues} proved that
\begin{equation*}
	\widehat{\lambda} - np \overset{\operatorname{d}}{\longrightarrow}
	\mathcal{N}(1-p,2p(1-p)) \qquad \text{~as~} n \rightarrow \infty.
\end{equation*}
More recently, \cite{fan2022asymptotic} established that $|\langle \widehat{\mathbf{u}}, \mathbf{u}\rangle|^{2}$ converges in distribution to a standard normal random variable as $n \rightarrow \infty$ after appropriate centering and scaling. However, the correlation structure between $\widehat{\lambda}$ and $\widehat{\mathbf{u}}$ at the scaling of asymptotic normality is not fully understood. Furthermore, and importantly, $\langle \mathbf{A} \boldsymbol{1}, \boldsymbol{1} \rangle$ and $\widehat{\lambda}$ are strongly correlated due to the fact that $\mathbf{A}$ is approximately a rank-one matrix. Taking all of these observations into account plausibly suggests that, under suitable conditions, $Q_{\Res} \overset{\operatorname{d}}{\approx} \mathcal{N}(\mu_{p}, \sigma_{p}^{2})$ for explicit expressions $\mu_{p}$ and $\sigma_{p}^{2}$, and more generally provided $K = O(1)$. Preliminary simulations appear to support this conjectured behavior. Crucially, here the modularity scaling is $O(1)$, in contrast to $O(\rho_{n}^{-1/2} n^{-1}) = O(n^{-1})$ for the likelihood and spectral variants in the article for the dense regime $\rho_{n} \equiv 1$.

\subsection{Extensions to more general latent space models}
\label{sec:suppl:beyond_sbm}

\begin{figure}[ht]
	\includegraphics[width=12cm, height=6cm]{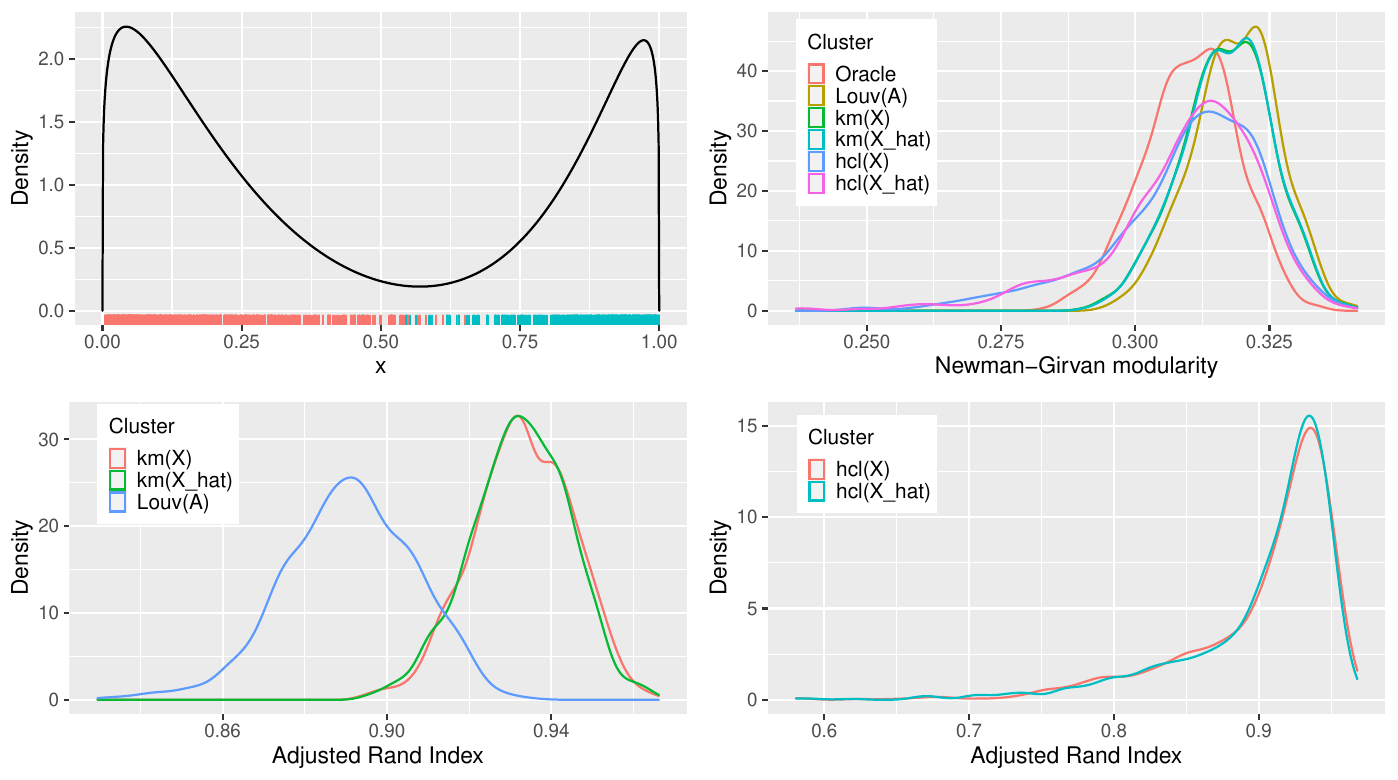}
	\centering
	\caption{Mild mixture distribution setup: $t_{i} \sim 0.6\times \operatorname{Beta}(1.2,5.5) + 0.4\times \operatorname{Beta}(8,1.2)$. Clock-wise from top-left: density plot with rug plot colored as per oracle cluster of underlying distribution; Density of $Q_{\Ng}/2m$ for different clusters; density of ARI between clusters from hierarchical clustering (for both $\mathbf{X}$ and $\widehat{\mathbf{X}}$) and oracle; density of ARI between clusters from $k$-means clustering (for both $\mathbf{X}$ and $\widehat{\mathbf{X}}$) and oracle as well as Louvain clustering and oracle.}
	\label{hardy_plt1}
\end{figure}

\begin{figure}[ht]
	\includegraphics[width=12cm, height=6cm]{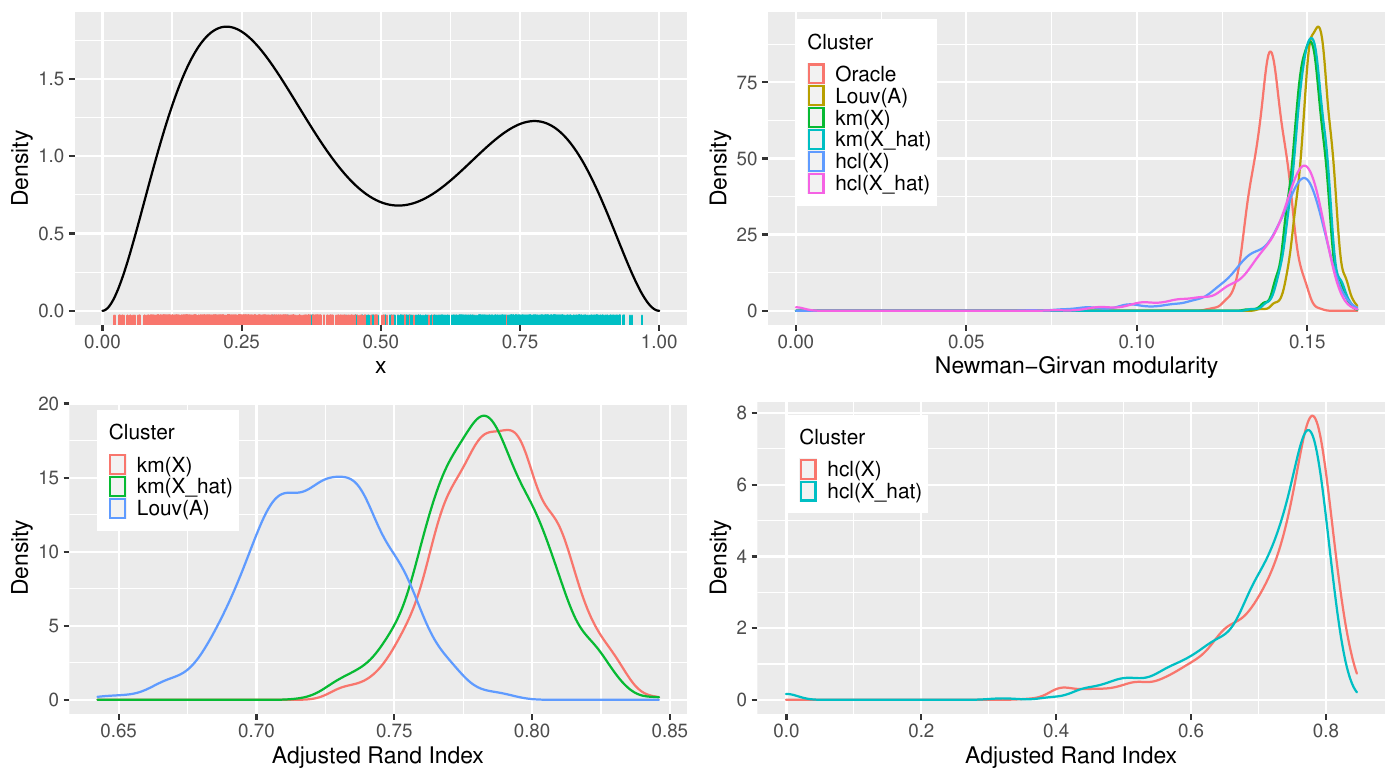}
	\centering
	\caption{Moderate mixture distribution setup: $t_{i} \sim 0.6\times \operatorname{Beta}(3,8) + 0.4\times \operatorname{Beta}(8,3)$. Clock-wise from top-left: density plot with rug plot colored as per oracle cluster of underlying distribution; Density of $Q_{\Ng}/2m$ for different clusters; density of ARI between clusters from hierarchical clustering (for both $\mathbf{X}$ and $\widehat{\mathbf{X}}$) and oracle; density of ARI between clusters from $k$-means clustering (for both $\mathbf{X}$ and $\widehat{\mathbf{X}}$) and oracle as well as Louvain clustering and oracle.}
	\label{hardy_plt2}
\end{figure}

This section briefly discusses possible extensions to more expressive latent space models, namely generalized random dot product graphs (GRDPGs) from \cite{rubindelanchy2022astatistical} in which edges are generated independently as

\begin{equation*}
    A_{ij} \mid \mathbf{X} \sim \operatorname{Bernoulli}\left((\mathbf{X} \mathbf{I}_{p,q} \mathbf{X}^{\top})_{ij}\right),\quad i \le j,
\end{equation*}
where $p \ge 1$, $q \ge 0$ are integers such that $p+q = d$ with $\mathbf{I}_{p,q} = \mathbf{I}_{p} \oplus (-\mathbf{I}_{q})$, and where the rows of the $n \times d$ matrix $\mathbf{X}$ are vertex-specific so-called latent position vectors. For GRDPGs, which are capable of exhibiting richer latent space geometry than block structure, care must be taken when seeking to relate network connectivity properties with latent vector properties and the concept of network modularity. For ease of discussion, we focus on the case when $q=0$, namely the consideration of so-called random dot product graphs (RDPGs).

Stochastic blockmodels are special cases of GRDPGs, and similarly, SBMs with positive semidefinite connectivity matrices are special cases of RDPGs. In particular, in a $K$-block SBM, the rows of the latent position matrix $\mathbf{X}$ each take one of $K$ distinct values based on the block membership of the corresponding node. In this case, the community membership vector $\boldsymbol{\tau}$ is either predefined with $\tau_{i} = \tau_{j}$ if and only if $\mathbf{X}_{i\cdot} = \mathbf{X}_{j\cdot}$, namely the $i$-th and $j$-th row vectors of $\mathbf{X}$ are equal. In general, an explicit concept or definition of community membership is needed in order to define a modularity value. One possible approach for RDPGs is to assign $\tau_{i} = \tau_{j}$ if and only if $d(\mathbf{X}_{i\cdot}, \mathbf{X}_{j\cdot}) \le \epsilon$ for some specified distance metric $d(\cdot,\cdot)$ and user-selected tolerance $\epsilon>0$.

The example provided here is inspired by the treatment in \cite{athreya2021estimation}. First, generate i.i.d. random variables $t_{i} \sim p \times \operatorname{Beta}(a_{1},b_{1}) + (1-p) \times \operatorname{Beta}(a_{2},b_{2})$, where $0 < p < 1$. Second, determine the latent position vectors via the function $\mathbf{x}(t) \coloneqq [t^{2},2t(1-t),(1-t)^{2}]$, yielding points (vectors) on the so-called one-dimensional Hardy--Weinberg curve in the unit simplex. Given $i \in \llbracket n \rrbracket$, the latent positions $\mathbf{x}(t_{i})$ are aggregated row-wise to form the matrix $\mathbf{X}$. In this example, an ``oracle" cluster membership is available via the coefficients of the mixture distribution. We choose $d(\mathbf{x}, \mathbf{y}) \coloneqq \|\mathbf{x} - \mathbf{y}\|_{\ell_{2}}$ and opt to hard-cluster the latent positions as our proxy for ground truth. In general, supposing the underlying mixture were composed of $K$ distributions, one could choose the number of true clusters to be larger or smaller than $K$ depending on the separability of the component distributions.

We consider two modeling situations. In \cref{hardy_plt1}, $t_{i} \sim 0.6\times \operatorname{Beta}(1.2,5.5) + 0.4\times \operatorname{Beta}(8,1.2)$ exhibits mild overlap between the component distributions. In \cref{hardy_plt2}, $t_{i} \sim 0.6\times \operatorname{Beta}(3,8) + 0.4\times \operatorname{Beta}(8,3)$ exhibits moderate overlap between the component distributions. The networks generated subsequently have $n=2000$ vertices and are independently simulated $1000$ times. Hierarchical clustering with average linkage on $\mathbf{X}$ is chosen to partition the nodes into two communities. For comparison, the same technique is applied to $\widehat{\mathbf{X}}$, the adjacency spectral embedding of $\mathbf{A}$. We also use the $k$-means algorithm on both $\mathbf{X}$ and $\widehat{\mathbf{X}}$ since the hard clustering technique employs the $\ell_{2}$ distance. Further, we apply the Louvain algorithm as a choice for community detection that directly takes the networks as input. In the two figures, we show the underlying mixture distributions, the Newman--Girvan modularity for the different sources of clusters, $Q_{\Ng}/2m$, and the ARI between the oracle clusters and each of the different true and estimated clusterings. The ARI plots show that in the mild mixture setup, estimated clustering with $\widehat{\mathbf{X}}$ captures the corresponding true clustering with high accuracy. In contrast, for the moderate mixture, deteriorated performance is observed. Louvain clustering performs much worse in recovering the oracle in both examples. The density plots for $Q_{\Ng}/2m$ repeatedly exhibit relatively large modularity values in the mild mixture case for all sources of clusters. 

The Newman--Girvan modularity, $Q_{\Ng}$, is the most commonly used modularity function for partitioning networks. The underlying null network depends only on the degree distribution of the observed graph. For one-dimensional RDPGs, consider $\mathbf{X} = [k_{1}/\sqrt{2m},k_{2}/\sqrt{2m},\dots,k_{n}/\sqrt{2m}]^{\top}$. In such cases, clustering based on $\mathbf{X}$ will resemble clustering based on node-degrees. In certain cases, the Newman--Girvan modularity shows relatively poor performance in detecting node-clusters and fails in some other situations as shown in \cref{sec:simulations}. One might instead select $Q_{\Like}$ as a choice for modularity if the edge generating distribution is plausibly Bernoulli. If, as for SBMs, the adjacency matrix $\mathbf{A}$ is generated using a block probability matrix $\mathbf{B}$ where the underlying community structure is unknown, then the $(A_{ij}-B_{\tau_{i}\tau_{j}})$ can be interpreted as mean zero noise. If the objective is to detect clusters via noise or residual minimization, one may consider the least squares approach in \cite{lei2019consistent}.

\subsection{Power analysis}
\label{sec:suppl:power}

This section provides a brief illustration pertaining to the discussion in \cref{sec:hypothesis_test}. Consider $\mathbf{B}^{(0)} = \mathbf{v}\mathbf{v}^{\top}$ and $\mathbf{B}^{(1)} = \mathbf{u}\mathbf{u}^{\top}$ where $u_{1} = v_{1}+\epsilon$, $u_{2} = v_{2}$. Setting $\mathbf{v} = \left[3/4,1/4\right]^{\top}$, $\boldsymbol{\pi} = [1/4,3/4]^{\top}$ and $\rho_{n} \equiv 1$, we generate $\mathbf{A} \sim \operatorname{SBM}(\mathbf{B}^{(1)},\boldsymbol{\pi})$ for $n \in \{100,500,1000\}$ and $\epsilon \in \{0.02,0.05\}$. We report the empirical performance across $1000$ independent simulation trials. \cref{tab_power_eps002} presents results for $\epsilon = 0.02$, while \cref{tab_power_eps005} presents results for $\epsilon = 0.05$. Both are as expected. Here, the connectivity matrix under the null is
\begin{equation*}
    \mathbf{B}^{(0)}
    =
    \begin{bmatrix}
	0.5625 & 0.1875\\
	0.1875 & 0.0625
	\end{bmatrix},
\end{equation*}
whereas the choices $\epsilon = 0.02$ and $0.05$ yield alternatives
\begin{equation*}
	\mathbf{B}^{(1)} = \begin{bmatrix}
	0.5929 & 0.1925\\
	0.1925 & 0.0625
	\end{bmatrix} \; \text{and} \, 
	\begin{bmatrix}
	0.6400 & 0.2000\\
	0.2000 & 0.0625
	\end{bmatrix},
	\text{~respectively.}
\end{equation*}

\begin{table}[htbp]
    \begin{center}
        \begin{tabular}{| c | c | c | c | c | c | c | }
        \hline
        $n$
            & \multicolumn{3}{c|}{Power for $\mathrm{T}_{\Like}$} & \multicolumn{3}{c|}{Power for $\mathrm{T}_{\Spec}$}
            \\
            \cline{2-7}
               &   Analytic  &  Empirical & Emp.~with plug-in
               &   Analytic  &  Empirical & Emp.~with plug-in
            \\
            \hline
            $100$ &  $0.093$  & $0.099$  & $0.094$  &  $0.1284$ & $0.152$ & $0.147$
            \\
            \hline
            $500$ &  $0.8641$  & $0.855$  & $0.856$  &  $0.9772$ & $0.938$ & $0.938$
            \\
            \hline
            $1000$ &  $0.99998$  & $1$  & $1$  &  $1$ & $1$ & $1$
            \\
            \hline
        \end{tabular}
    \caption{Power analysis for $\mathrm{T}_{\Like}$ and $\mathrm{T}_{\Spec}$ when $\epsilon = 0.02$.}
    \label{tab_power_eps002}
    \end{center}
\end{table}
\begin{table}[htbp]
    \begin{center}
        \begin{tabular}{| c | c | c | c | c | c | c | }
        \hline
        $n$
            & \multicolumn{3}{c|}{Power for $\mathrm{T}_{\Like}$} & \multicolumn{3}{c|}{Power for $\mathrm{T}_{\Spec}$}
            \\
            \cline{2-7}
               &   Analytic  &  Empirical & Emp.~with plug-in
               &   Analytic  &  Empirical & Emp.~with plug-in
            \\
            \hline
            $100$ &  $0.3424$  & $0.357$  & $0.365$  &  $0.5426$ & $0.549$ & $0.555$
            \\
            \hline
            $500$ &  $1$  & $1$  & $1$  &  $1$ & $1$ & $1$
            \\
            \hline
            $1000$ &  $1$  & $1$  & $1$  &  $1$ & $1$ & $1$
            \\
            \hline
        \end{tabular}
    \caption{Power analysis for $\mathrm{T}_{\Like}$ and $\mathrm{T}_{\Spec}$ when $\epsilon = 0.05$.}
    \label{tab_power_eps005}
    \end{center}
\end{table}

\subsection{Example~4:~graphs with $K=3$, $d=1$, balanced, core-periphery}
\label{eg4}
\begin{figure}[ht]
	\includegraphics[width=10cm, height=8cm]{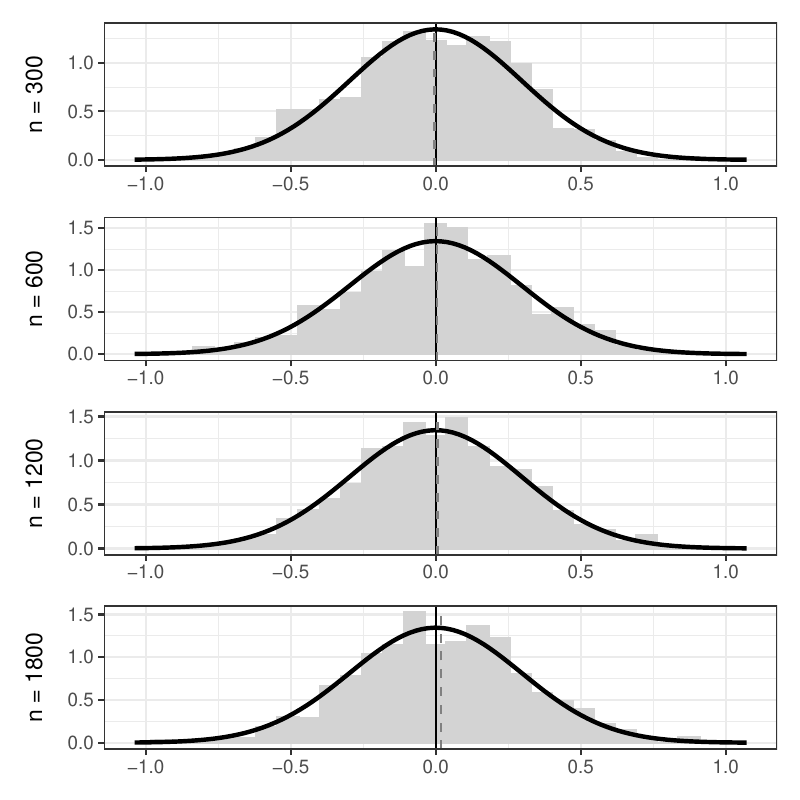}
	\centering
	\caption{Dense networks and residual-based modularity in \cref{eg4}. Dashed vertical line shows bias in simulation. Solid vertical line shows population bias. Solid curve shows population density fit.}
	\label{eg4_resid}
\end{figure}

Consider a three-block SBM where $\mathbf{B}$ takes the form
$\left[\begin{smallmatrix}
  p\\
  p^{2}\\
  p^{3}
\end{smallmatrix}\right] \times 
\left[\begin{smallmatrix} p &  p^{2} & p^{3} \end{smallmatrix} \right]$.
Set 
\begin{equation*}
    p=3/4, \quad
    \mathbf{B} = \begin{bmatrix}
                    0.5625000 & 0.4218750 & 0.3164062\\
                    0.4218750 & 0.3164062 & 0.2373047\\
                    0.3164062 & 0.2373047 & 0.1779785
                \end{bmatrix},
    \quad
    \boldsymbol{\pi}=\begin{bmatrix}
                    1/3\\
                    1/3\\
                    1/3
                \end{bmatrix},
    \quad
    n \in \{300,600,1200,1800\}.
\end{equation*}
We again simulate $1000$ independent trials. The results for residual-based modularity are presented in \cref{eg4_resid}. The Louvain algorithm shows similar performance as in the earlier example; ARI is almost always near zero, and the number of clusters is dramatically overestimated.

\subsection{Further examples illustrating modularity asymptotics}
\label{sec:suppl:further_eg}
\begin{figure}[ht]
	\includegraphics[width=14cm, height=8cm]{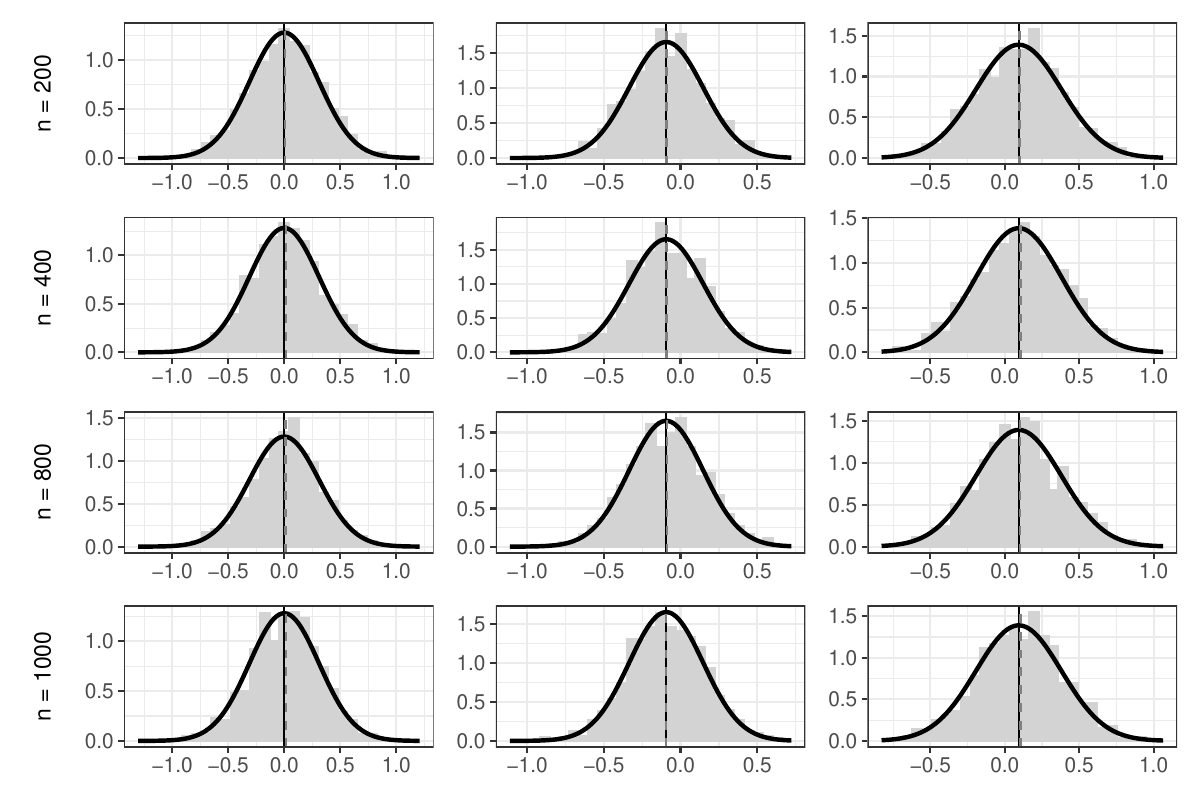}
	\centering
	\caption{Dense case simulations for \cref{eg3}. Left to right panels are for $\rho_{n}^{-1/2}n^{-1}Q_{\Like}$, $\rho_{n}^{-1/2}n^{-1}Q_{\Spec}$, and $\rho_{n}^{-1/2}n^{-1}Q_{\Res}$, respectively. Dashed vertical line shows bias in simulation. Solid vertical line shows population bias. Solid curve shows population density fit.}
	\label{eg3_dense}
\end{figure}

\cref{sec:simulations} illustrates theoretical properties of the modularity functions via simulation. In \cref{eg3,eg4} which consider rank-degenerate $\mathbf{B}$ matrices, the residual-based modularity function is investigated. For completeness, here we show the performance of $Q_{\Like}$ and $Q_{\Spec}$ in each of these settings. Further, for the setup in \cref{eg3} we now present the $\rho_{n} \equiv 1$ setting in \cref{eg3_dense}. The empirical values of the parameters closely agree with the asymptotic, analytic population values. Here, too, the performance of the Louvain algorithm is predictably poor, with adjusted Rand index values near zero and estimated number of clusters much larger than the ground truth value, since the underlying block probability matrix has core-periphery structure.

\begin{figure}[ht]
	\includegraphics[width=14cm, height=8cm]{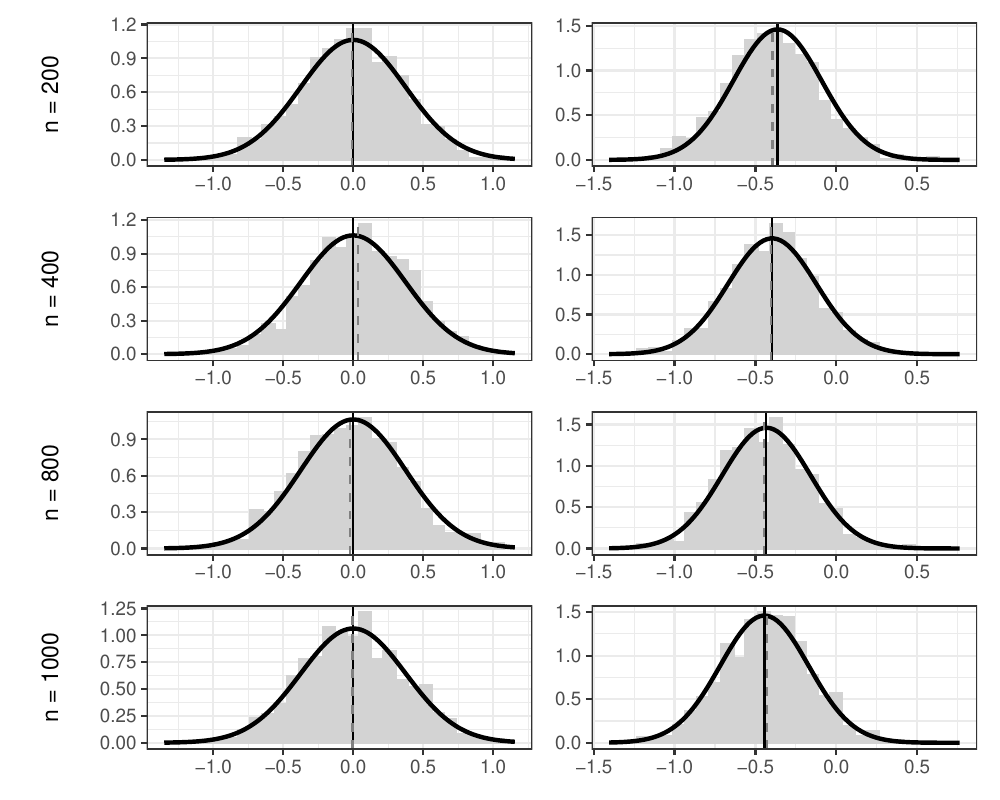}
	\centering
	\caption{Sparse case simulations for \cref{eg3}. Left panel is $\rho_{n}^{-1/2}n^{-1}Q_{\Like}$ and right panel is $\rho_{n}^{-1/2}n^{-1}Q_{\Spec}$. Dashed vertical line shows bias in simulation. Solid vertical line shows population bias. Solid curve shows population density fit.}
	\label{eg3_sparse}
\end{figure}

\begin{figure}[ht]
	\includegraphics[width=14cm, height=8cm]{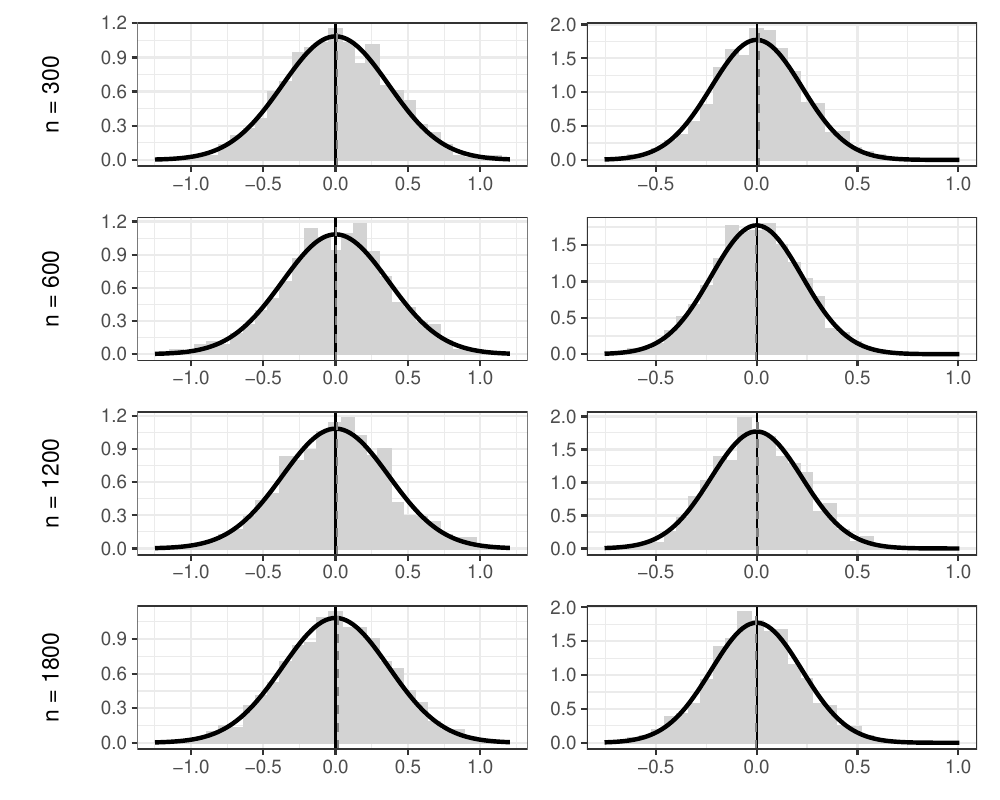}
	\centering
	\caption{Additional simulations for \cref{eg4}. Left and right panels are for $\rho_{n}^{-1/2}n^{-1}Q_{\Like}$ and $\rho_{n}^{-1/2}n^{-1}Q_{\Spec}$, respectively. Dashed vertical line shows bias in simulation. Solid vertical line shows population bias. Solid curve shows population density fit.}
	\label{eg4_dense}
\end{figure}

\subsection{COBRE data: additional discussion}
\label{sec:suppl:extra_COBRE}

The Power parcellation consists of fourteen brain systems: 1.~\emph{Sensory/somatomotor Hand} (30 nodes), 2.~\emph{Sensory/somatomotor Mouth} (5 nodes), 3.~\emph{Cingulo-opercular Task Control} (14 nodes), 4.~\emph{Auditory} (13 nodes), 5.~\emph{Default mode} (58 nodes), 6.~\emph{Memory retrieval} (5 nodes), 7.~\emph{Visual} (31 nodes), 8.~\emph{Fronto-parietal Task Control} (25 nodes), 9.~\emph{Salience} (18 nodes), 10.~\emph{Subcortical} (13 nodes), 11.~\emph{Ventral attention} (9 nodes), 12.~\emph{Dorsal attention} (11 nodes), 13.~\emph{Cerebellar} (4 nodes), and 14.~\emph{Uncertain} (28 nodes). On the basis of function, we also consider a coarser partition into $\widehat{K}=5$ communities, specifically of the form $\{\{1,2,3,4\}, \{5,6,8,9\}, \{7,12\}, \{10,11,13\}, \{14\}\}$.

\end{document}